\documentclass[a4paper,11pt]{article}
\pdfoutput=1 % if your are submitting a pdflatex (i.e. if you have
\usepackage{jheppub} % for details on the use of the package, please
\usepackage[T1]{fontenc}

\usepackage{amsfonts}
\usepackage{float}
\usepackage{placeins}
\usepackage{graphicx}

\usepackage{color,amsmath,amssymb,bm}

\def\blfootnote{\xdef\@thefnmark{}\@footnotetext}
\def\eg{{\em e.g.}}
\def\ie{{\em i.e.}}

\newcommand{\qhat}{\hat{q}}
\newcommand{\Tpc}{$T_{\rm pc}$}
\newcommand{\beq}{\begin{equation}}
\newcommand{\eeq}{\end{equation}}
\newcommand{\bea}{\begin{eqnarray}}
\newcommand{\eea}{\end{eqnarray}}
\newcommand{\ltsim}{\raisebox{-4pt}{$\,\stackrel{\textstyle <}{\sim}\,$}}

\newcommand{\add}[1]{{\bf\textcolor{blue}{#1}}}

\title{Nonperturbative Effects on Radiative Energy Loss of Heavy Quarks}
\author[a,b,c,1]{Shuai Y.F.~Liu, \note{Corresponding author.}}
\author[a]{and Ralf Rapp}

\affiliation[a]{Cyclotron Institute and Department of
Physics and Astronomy, Texas A\&M University, College Station, TX 77843-3366, USA}
\affiliation[b]{Quark Matter Research Center, Institute of Modern Physics,	Chinese Academy of Sciences, Lanzhou, Gansu, 730000, China}
\affiliation[c]{University of Chinese Academy of Sciences, Beijing, 100049, China}

\emailAdd{lshphy@gmail.com}
\emailAdd{rapp@comp.tamu.edu}

\date{\today}
\abstract{
The radiative energy loss of fast partons traveling through the quark-gluon plasma (QGP) is commonly studied within perturbative 
QCD (pQCD). Nonperturbative (NP) effects, which are expected to become important near the critical temperature, have been 
much less investigated. Here, we utilize a recently developed $T$-matrix approach to incorporate NP effects for gluon emission 
off heavy quarks propagating through the QGP. We set up four cases that contain, starting from a Born diagram calculation 
with color-Coulomb interaction, an increasing level of NP components, by subsequently including (remnants of) confining 
interactions, resummation in the heavy-light scattering amplitude, and off-shell spectral functions for both heavy and light 
partons. For each case we compute the power spectra of the emitted gluons, heavy-quark transport coefficients (drag and 
transverse-momentum broadening, $\qhat$), and the path-length dependent energy loss within a "QGP brick" at fixed temperature. 
Investigating the differences in these quantities between the four cases illustrates how NP mechanisms affect gluon radiation 
processes. While the baseline perturbative processes experience a strong suppression of soft radiation due to thermal 
masses of the emitted gluons, confining interactions, ladder resummations and broad spectral functions  (re-)generate a large enhancement toward low 
momenta and low temperatures. For example, for a 10 GeV charm quark at 200 MeV temperature, they enhance the transport 
coefficients by up to a factor of 10, while the results smoothly converge to perturbative results at sufficiently hard scales.
}

\begin{document}
%\pacs{25.75.Dw, 12.38.Mh, 25.75.Nq}
%\keywords{Heavy-Flavor Transport, Quark-Gluon Plasma, Ultrarelativistic Heavy-Ion Collisions}
\maketitle
\flushbottom
%%%%%%%%%%%%%%%%%%%%%%%%%%%%%%%%%%%%%%%%%%%%%%%%%%%%%%%%
\section{Introduction}
\label{sec_intro}
%%%%%%%%%%%%%%%%%%%%%%%%%%%%%%%%%%%%%%%%%%%%%%%%%%%%%%%
In ultra-relativistic heavy-ion collisions (URHICs) heavy quarks are mostly produced in primordial hard collisions of the incoming 
nucleons, on a short timescale, $\tau_{\rm prod}\sim 1/2m_Q$, governed by the heavy-quark (HQ) mass, $m_Q$; therefore, 
they probe the entire evolution of the fireball formed in these reactions. The large mass of heavy quarks also enables theoretical 
simplifications in the description of their transport through the medium~\cite{Prino:2016cni}, allowing for rather direct 
connections between their microscopic interactions and pertinent observables in experiment. Heavy-flavor (HF) hadrons 
are therefore excellent probes of the properties of the QGP and its hadronization in 
URHICs~\cite{Svetitsky:1987gq,Moore:2004tg,vanHees:2005wb,Prino:2016cni,Dong:2019byy}.

One generally distinguishes two processes that control the dynamics of heavy quarks in the QGP -- collisional and radiative 
ones. The former are mainly responsible for thermalization of heavy quarks at low momenta, $p_Q\simeq \sqrt{3M_QT}$, 
while the latter become increasingly relevant at higher HQ momenta~\cite{Mustafa:2004dr,Gossiaux:2010yx,Cao:2013ita}.
Radiative processes, \ie,  medium-induced gluon emission, in the relativistic limit have been widely studied over the past two 
decades. Various formalisms, such as AMY~\cite{Arnold:2002ja}, ASW~\cite{Armesto:2003jh}, 
BDMPS-Z~\cite{Baier:1996kr,Zakharov:1996fv},  
DGLV~\cite{Gyulassy:2000er,Djordjevic:2004nq}, higher-twist~\cite{Wang:2001ifa,Majumder:2009ge}, and
%Gyulassy:1993hr,Wiedemann:2000za,Ovanesyan:2011xy} 
SCET~\cite{Ovanesyan:2011xy,Kang:2017frl},  have been developed for the phenomenology of light-flavor jet quenching in 
URHICs. Among them, some have been extended to massive 
quarks~\cite{Armesto:2003jh,Zhang:2003wk,Djordjevic:2009cr,Abir:2015hta,Kang:2016ofv,Du:2018yuf} 
and implemented into numerical simulations for their transport in 
URHICs~\cite{Uphoff:2014hza,Gossiaux:2010yx,Das:2010tj,Xu:2015bbz,Cao:2017hhk,Ke:2018tsh,Katz:2019fkc}, 
see also Refs.~\cite{Burke:2013yra} and \cite{Rapp:2018qla,Cao:2018ews} for recent comparisons between 
different energy loss formalisms for light-flavor jets and heavy quarks, respectively. Gluon emission within these approaches is generally described using perturbative-QCD (pQCD) methods. Within different expansion schemes, these approaches concentrate 
on the regions of the phase space where nonperturbative (NP) effects are expected to be small or can be absorbed into a 
transport parameter that encodes the medium properties, most notably the jet transport coefficient, $\hat{q}$ . So far, 
few studies of NP effects in radiative processes have been carried out at a microscopic level (see, \eg, 
Ref.~\cite{Xu:2015bbz}), especially in the relatively 
low-momentum and temperature regimes where standard expansion and factorization schemes are no longer reliable. 
The objective of the present work is to utilize a NP $T$-matrix approach to provide insights into mechanisms of 
radiative energy loss of heavy quarks which are not easily assessed in pQCD approaches.

The phenomenology of HF observables in URHICs, especially the measured elliptic flow, requires heavy quarks to 
have an interaction strength with the medium that goes well beyond pQCD~\cite{Moore:2004tg,vanHees:2004gq,Rapp:2018qla}, especially at low momentum and temperature. As rigorous QCD calculations are challenging in this regime, 
one has to resort to model approaches. An essential step in improving their reliability is to root the model parameters in 
fits to lattice-QCD (lQCD) data as much as possible. In our previous work we have developed a quantum many-body approach 
with a QCD-inspired model Hamiltonian~\cite{Liu:2016ysz,Liu:2017qah,Liu:2018syc} that turns out to be an effective 
tool in describing a wide variety of QGP properties with a relatively small number of parameters, including its equation of state 
(EoS), HQ free energies, and correlation functions, while predicting spectral properties and transport coefficients (\eg, shear 
viscosity and HQ diffusion coefficient). Key features of this approach are the inclusion of remnants of the confining force in the 
QGP, ladder resummed amplitudes leading to the dynamical generation of bound states as the pseudo-critical temperature, \Tpc, is approached from above, and a full off-shell treatment of propagators and scattering amplitudes which is, in fact, mandated by the large widths that develop for low-momentum partons. In this work, we will deploy this approach to calculate the radiative energy loss of heavy quarks. The off-shell 
transport theory underlying this framework is based on the Kadanoff-Baym equation following the same logic as used in
our previous study of the HQ collisional energy loss~\cite{Liu:2018syc}. 
%We start with the Kadanoff Baym used for Collisional process~\cite{Liu:2018syc} where the radiative process can be naturally added and then reduced to a classical Boltzmann equation that latter been reduced to Langevin equation.
To illustrate how NP physics affects gluon radiation, we set up a systematic comparison between four different scenarios. 
Case (1) is the perturbative baseline in our approach that only includes screened Coulomb interactions at the Born level. 
We add confining interactions in case (2) and the resummation of $t$-channel diagrams in case (3). In case (4), we further  
include the off-shell effects for parton and HQ propagators, \ie, their broad spectral functions, representing the full results 
in our current framework. We analyze the differences between these four setups for several quantities, such as emission 
power spectra, transport coefficients, and fractional energy loss within a QGP brick medium. 
%Understanding the systematic uncertainties between these different setups within the same model allows us to gain generic insights into the role non-perturbative effects on heavy quark energy loss. %can help us to cancel out some model artificial and extract the model independent insights.

The paper is organized as follows. In Sec.~\ref{sec_theo} we lay out our formalism for calculating the HQ radiative energy loss
within the $T$-matrix formalism (Sec.~\ref{ssec_form}), disucss its realtion to pQCD diagrams (sec.~\ref{ssec_diag}), and 
define four different model setups with an increasing content of NP components (Sec.~\ref{ssec_setup}). 
In Sec.~\ref{sec_num}, we compare the numerical results for typical energy loss quantities from these four cases. A summary and outlook is given in Sec.~\ref{sec_con}.

%%%%%%%%%%%%%%%%%%%%%%%%%%%%%%%%%%%%%%%%%%%%%%%%%%%%%%%
\section{Gluon Radiation in the $T$-Matrix Formalism}
\label{sec_theo}
%%%%%%%%%%%%%%%%%%%%%%%%%%%%%%%%%%%%%%%%%%%%%%%%%%%%%%%
In this section, we first develop the formalism for computing the radiative energy loss in the context of the in-medium $T$-matrix 
(Sec.~\ref{ssec_form}), discuss the relation of the NP calculation to the diagrams in pQCD including the issue of gauge invariance (Sec.~\ref{ssec_diag}), and then introduce four different 
model cases of increasing levels of NP effects to illustrate how these influence radiative processes of charm quarks (Sec.~\ref{ssec_setup}).

%%%%%%%%%%%%%%%%%%%%%%%%%%%%%%%%%
\subsection{Many-body formalism of heavy-quark radiative energy loss}
\label{ssec_form} 
%%%%%%%%%%%%%%%%%%%%%%%%%%%%%%%%%%
The derivation of the formalism for evaluating radiative processes is similar to that in our previous work~\cite{Liu:2018syc}. We 
first recall the calculation for collisional processes and then derive the equations for radiative processes.

We start with a Kadanoff-Baym equarion for the heavy quark ($Q$) in terms of its Green's function, $G_Q$ and selfenergy, 
$\Sigma_Q$, 
\begin{align}
\frac{\partial}{\partial t}[\int d\omega G_Q^<(\omega,\textbf{p},t)]&=
\int d\omega[i\Sigma_Q^{<}(\omega,\textbf{p},t)G_Q^>(\omega,\textbf{p},t)
- i\Sigma_Q^{>}(\omega,\textbf{p},t)) G_Q^<(\omega,\textbf{p},t)] \ ,
\label{KB} 
\end{align}
where the $>,<$ superscripts denote the fixed-ordered quantities in real-time many-body field theory (sometimes referred to as Wightman functions). These quantities
depend  on Wigner coordinates $\omega$ and $\bf{p}$ with their Fourier conjugates $t_1-t_2$ and $\bf{x}_1-\bf{x}_2$, 
respectively. By putting the incoming heavy quark on-shell~\cite{Liu:2018syc}, Eq.~(\ref{KB}) can be reduced to a semi-classical Boltzmann equation for its phase space distribution (PSD) function, $f_Q$
\begin{align}
\frac{\partial}{\partial t}f_Q(\textbf{p},t)&=
\int \frac{d^3\textbf{k}}{(2\pi)^3} 
[w(\textbf{p+k},\textbf{k})f_Q(\textbf{p+k},t)-w(\textbf{p},\textbf{k})f_Q(\textbf{p},t)] \  ,
\label{boltz} 
\end{align}
where the transition rate $ w(\textbf{p},\textbf{k})$ encodes the quantum many-body information of the system, which can be used to obtain transport coefficients~\cite{Svetitsky:1987gq}. The first and second terms on the right hand side of 
Eq.~(\ref{boltz}) correspond to the first and second terms of Eq.~(\ref{KB}) respectively, as discussed in Ref.~\cite{Liu:2018syc}.

For the collisional 2$\to$2 scattering processes, $ \Sigma_Q^{>} $ in Eq.~(\ref{KB}) can be schematically expressed via
a HQ scattering amplitude off thermal partons, $M_{Qi \to Qi}$, as
\begin{align}
\Sigma_Q^{>}(p)=\int d^{4}\tilde{p}'   d^{4}\tilde{q} d^{4}\tilde{q}' (2\pi)^4\delta^{(4)} |M_{Qi\leftrightarrow Qi}|^2G_Q^>(p')G_i^<G(q)_i^>(q') \  , 
\label{eq_sigmalarge} 
\end{align}
where $p$ ($p'$) and $q$ ($q'$) are the 4-momenta of the incoming (outgoing) heavy quark and thermal parton, $i$, respectively.  
We use the notation $ d^4\tilde{p}=d^4p/(2\pi)^3 2 \epsilon(p) $ for the 4-momentum phase space elements (with on-shell 
energy $\epsilon_Q(p)$)  and $\delta^{(4)}$ for the 4-momentum conserving Dirac delta function.
The  Green's functions can be represented by the spectral functions ($\rho$) with Bose/Fermi factors, $n_{i,Q}$, as
\begin{align}
&G_Q^{>}(\omega,p)=- i(2\pi)\rho_Q(\omega,p)(1- n_Q(\omega)) \ ,
\nonumber\\
&G_i^{<}(\omega,p)=\mp i(2\pi)\rho_i(\omega,p)n_i(\omega) \ ,
\nonumber\\
& G_i^{>}(\omega,p)=- i(2\pi)\rho_i(\omega,p)(1\pm n_i(\omega)) \ ,
\label{eq_approx}
\end{align}
The transition rate, $w(\textbf{p},\textbf{k})$, derived from $ \Sigma_Q^{>} $ can then be expressed as
\begin{align}
w(\textbf{p},\textbf{k})=&\int d^4\tilde{q}   d^4\tilde{q}' d\omega'(2\pi)^4\delta^{(4)} 
|M_{Qi\leftrightarrow Qi}|^2 
 \rho_Q \left[1- n_Q\right] \rho_i n_i \rho_i \left[1\mp n_i\right]  \ ,
\label{eq_wrate}
\end{align}
where $ \omega' $ is the energy of the outgoing heavy quark.
The off-shell spectral functions of both thermal partons and heavy quarks ($x=Q,i$), which encode the non-trivial quantum many-body physics of the system, 
can be expressed as
%~\footnote{The radiative contribution to the spectral function is more subtle and be discuss separately latter}
\begin{align}
\rho_x(k)&=-\text{Im}\left(\frac{1}{\omega-\varepsilon_x(p)-\Sigma_x(\omega,p)}\right)\,,
\label{eq_spec}
\\
\text{Im}\Sigma_x(p)&=-\sum_{j,x}\int d^{4}\tilde{p}'   d^{4}\tilde{q} d^{4}\tilde{q}' (2\pi)^4\delta^4 |M_{xj\leftrightarrow xj}|^2 \rho_j(q)\rho_{j}(q')\rho_{x}(p')
\nonumber\\
 & \qquad \qquad \times  [n_j(1-n_{j})(1-n_{x})+(1-n_j)n_{x}n_{j}]\nonumber\\
&\equiv \sum_{j,x}\int d^{4}\tilde{q}~\text{Im}M_{xj\to xj}\,\rho_j(q) (n_j\pm n_{jx}) \  ,
\label{eq_selfeels}
\end{align}
where the $ M_{xj\to xj}$ are "elastic" heavy-light or light-light scattering amplitudes. The summation is over all internal degrees of freedom with spin ($s$), color ($c$) and flavor ($f$) degeneracy factors and averaging procedure. The second line of Eq.~(\ref{eq_selfeels}) is related to its first line through the optical theorem. We 
utilize the second line to calculate the elastic selfenergy as detailed in Ref.~\cite{Liu:2017qah}. The real part of the selfenergy 
is obtained from a dispersion relation.

%The drag coefficient is 
%\begin{align}
%A(\textbf{p})=&\int d^{12}\tilde{q}(2\pi)^4\delta^{(4)} 
%|M_{Qi\leftrightarrow Qi}|^2 
% \rho_Q \left[1- n_Q\right] \rho_i n_i \rho_i \left[1\mp n_i\right](1-\frac{\textbf{p}\cdot\textbf{p}'}{\textbf{p}^2})  \ ,
%\label{eq_wrate}
%\end{align}
%For the extension, we need to do for including the radiative energy loss is just to replace the collisional matrix element   $ M_{2\leftrightarrow2} $ in Eq.~(\ref{eq_sigmalarge}) and Eq.~(\ref{eq_wrate}) to radiative matrix element $ M_{1\leftrightarrow2} $ and remove the one in coming green function $ G_i^< $. Note the matrix amplitudes in spectral function evaluation are unchanged. Thus the equations 
The key step to extend the above formalism from collisional to radiative processes is to replace the collisional amplitude 
$ M_{2\to2} $  in Eq.~(\ref{eq_sigmalarge}) with radiative amplitude $ M_{1\to2} $ and remove 
one incoming Green's function, $ G_i^< $. This leads to
\begin{align}
&\Sigma_Q^{>}(p)=\int d^{4}\tilde{p}'  d^{4}\tilde{k}  (2\pi)^4\delta^4 |M_{Q\to Qg}|^2 G_Q^>G_g^> \ ,
\label{eq_radse}\\
&w(\textbf{p},\textbf{k})=\int d^4\tilde{p}' d\nu (2\pi)^4\delta^{(4)}|M_{Q\to Qg}|^2 
\rho_Q \left[1- n_Q\right] \rho_g \left[1+n_g\right]   \  ,
\label{eq_radrate}
%\\&A(\textbf{p})=\int d^{8}\tilde{q}(2\pi)^4\delta^{(4)} 
%|M_{Q\leftrightarrow Qg}|^2
%\rho_Q \left[1- n_Q\right]\rho_i \left[1\mp n_i\right](1-\frac{\textbf{p}\cdot\textbf{p}'}{\textbf{p}^2})  \ ,\label{eq_raddrag}
\end{align}
where $k=(\nu,\textbf{k})$ is he outgoing gluon  4-momentum.
The amplitude for the $Q\to Qg$ process is calculated using the pQCD diagrams for gluon emission, 
\begin{align}
&|M_{Q\to Qg}|^2\approx \sum_{c,s}g^2|\bar{u}(p')\gamma_\mu u(p) \epsilon^\mu(k)|^2=\nonumber\\
&4d_Q C_F g^2\left\{\left(\varepsilon_Q(p)\varepsilon_Q(p')-m_Q^2\right)-\frac{(\textbf{p}\cdot\textbf{k})(\textbf{p}'\cdot\textbf{k})}{\textbf{k}^2}\right\} \,.
\label{eq_vertex}
\end{align}
We choose the standard Dirac spinor ($u$) and the polarization tensor ($\epsilon^\mu$) in Coulomb gauge where the sum 
of the polarization tensors is $\sum_s\varepsilon_\mu\varepsilon_\nu=\delta_{ij}-k_ik_j/k^2$. The amplitude is evaluated using on shell energies where the in-medium HQ mass is 
determined by the many-body formalism~\cite{Liu:2018syc}, while, for reasons of gauge invariance, we restrict the gluon polarization tensor to its transverse components, \ie,  $ A_{T}=(\delta_{ij}-k_ik_j/k^2)A_j$ is invariant under the gauge transformation with parameter $\alpha$ as $A_{T}'=(\delta_{ij}-k_ik_j/k^2)(A_j +\alpha k_j)= A_{T}$. This setup approximations neglects other gauge invariant contributions, in particular vertex corrections which are, however, rather involved and beyond the scope of the present investigation. Note that in their in-medium spectral functions figuring in 
Eq.~(\ref{eq_spec}), the gluons have a finite mass (as dictated by requiring to reproduce the QGP EoS).

Inserting the amplitude of Eq.~(\ref{eq_vertex}) into Eq.~(\ref{eq_radrate}), we have
\begin{align}
w(\textbf{p},\textbf{k})\equiv&\frac{dN_g}{dtd^3\textbf{k}}\nonumber\\=&\frac{1}{2 \varepsilon_Q(p)}\int 
\frac{d\nu}{(2\pi)^3 2\varepsilon_g(k)}\frac{d\omega'd^3\textbf{p}' }{(2\pi)^3 2\varepsilon_Q(p')}
\delta(\varepsilon_Q(p)-\omega'-\nu)\delta^{(3)}(\textbf{p}-\textbf{p}'-\textbf{k})\frac{(2\pi)^4}{d_Q}\nonumber\\
&\times 4d_Q C_F g^2\left\{\left(\varepsilon_Q(p)\varepsilon_Q(p')-m_Q^2\right)-\frac{(\textbf{p}\cdot\textbf{k})(\textbf{p}'\cdot\textbf{k})}{\textbf{k}^2}\right\}\nonumber\\
&\times \rho_Q(\omega',p')\rho_i(\nu,k)[1-n_Q(\omega')][1+n_g(\nu')]\ .
\label{eq_ratedetail}
\end{align}
In leading order for the bare $Q\to Qg$ splitting process, a restriction to on-shell energies, $ \omega'=\varepsilon_Q(p') $ and $ \nu=\varepsilon_g(k) $ in the spectral functions leads to a vanishing rate $ w(\textbf{p},\textbf{k}) $, since the $\delta$-functions for energy and momentum conservation cannot be simultaneously satisfied. However, at the leading dressed order (skeleton order), 
$ w(\textbf{p},\textbf{k}) $ is finite since the spectral functions for outgoing heavy quark and gluons allow for off-shell energies, opening up phase space that simultaneously satisfies energy and momentum conservation. As shown in Eqs.~(\ref{eq_spec}) and 
(\ref{eq_selfeels}), the information on elastic scatterings between the outgoing partons ($ Q,g $) and the medium, and the mean 
free path of the outgoing partons, are encoded in the off-shell properties of the spectral functions.

%for dressed Green functions, the self-energies makes the outgoing heavy quarks and gluons off-shell which open up forbidden phase space so that make the rate is no-zero at leading (dressed) order. 

Following Ref.~\cite{Svetitsky:1987gq}, the transition rate $ w(\textbf{p},\textbf{k}) $ can be used to evaluate the drag coefficient (as used in a Langevin simulation below),  
\begin{align}
%I(p,k_T,k_L)=w(p,k_T,k_L)\sqrt{k_T^2+k_L^2},\qquad 
A(p)=\int d^3\textbf{k} w(\textbf{p},\textbf{k})\frac{\textbf{p}\cdot\textbf{k}}{\textbf{p}^2}= \int d^3\textbf{k}w(\textbf{p},\textbf{k})\frac{k_L}{p} \  ,
\label{eq_offtrans}
\end{align}
where $ k_L\equiv(\textbf{k}\cdot \textbf{p}/p) $ is the longitudinal momentum transfer.
For high-energy (or large $k$) radiative processes, $ k_L\approx k $ is a good approximation. Thus, it is useful to define the power 
function  $ k\, w(\textbf{p},\textbf{k}) $ so that the drag coefficient can be approximately expressed as 
$ A(p)\approx\int d^3\textbf{k} k\, w(\textbf{p},\textbf{k})/p$. If we further assume an azimuthal symmetry of the emission, 
we can define the power function per $k_T$ and $k_L$ phase space as
\begin{align}
W(p,k_L,k_T)\equiv k\frac{dN_g}{dt dk_Tdk_L}=(2\pi k_T)\sqrt{k_T^2+k_L^2} w(p,k_T,k_L) \ .
\label{eq_power}
\end{align}
The factor $2\pi k_T $ originates from integrating over the azimuth angle. This power function characterizes 
how the radiated energy is distributed over the phase space. We can integrate over $k_T$ to obain
\begin{align}
x \frac{dN_g}{dt dx}\approx(k/p)\frac{dN_g}{dt d(k_L/p)} =\int d k_T (2\pi k_T)\sqrt{k_T^2+k_L^2} w(p,k_T,k_L),
\label{eq_xdnddx}
\end{align}
where $ x=(k_L+\varepsilon_g(k))/(p+\varepsilon_Q(p))) $ ($ x\approx k_L/p $ at high momentum) is the longitudinal momentum fraction of the emitted gluon taken from the parent heavy quark.  The variable $x$ is often referred to as a light-front coordinate and commonly used in existing literature for radiative energy loss.

\subsection{Relation to pQCD diagrams}
\label{ssec_diag}
%%%%%%%%%%%%%%%%%%%%%%%%%%%%%%%

%The formalism for radiative gluon emission presented above, which is motivated from the many-body theory in the HQ limit, is not rigorously gauge invariant. Using a physical Coulomb gauge for the polarization tensor can help reduce gauge artifacts. In addition, 
%we have checked that using an alternate gauge for the polarization tensor -- the light-front gauge -- yields differences in the quantities of interest between these two gauges of order 10\%. Although this lends some support to the robustness of
%our results, it is important to understand why the formalism is not gauge invariant and what the potential ways are to improve 
%it in the future.\\ 
%{\red{It would be useful to show a figure to illustrate the gauge dependence}}.

To obtain a better understanding of how the NP framework employed here relates to pQCD calculations, we carry out a comparison of relevant diagrams in this section. This will also allow us to address the issue of gauge invariance for the NP case, which is a rather challenging one in the presence of resummed interactions.

\begin{figure} [th]
	\centering
	\includegraphics[width=1\columnwidth]{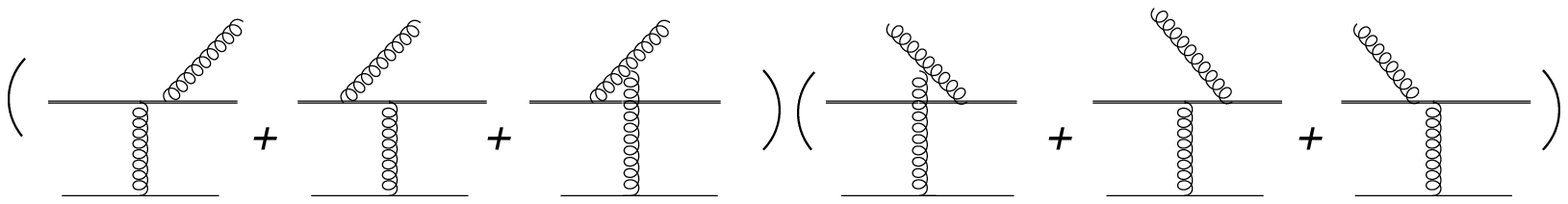}
	\includegraphics[width=1\columnwidth]{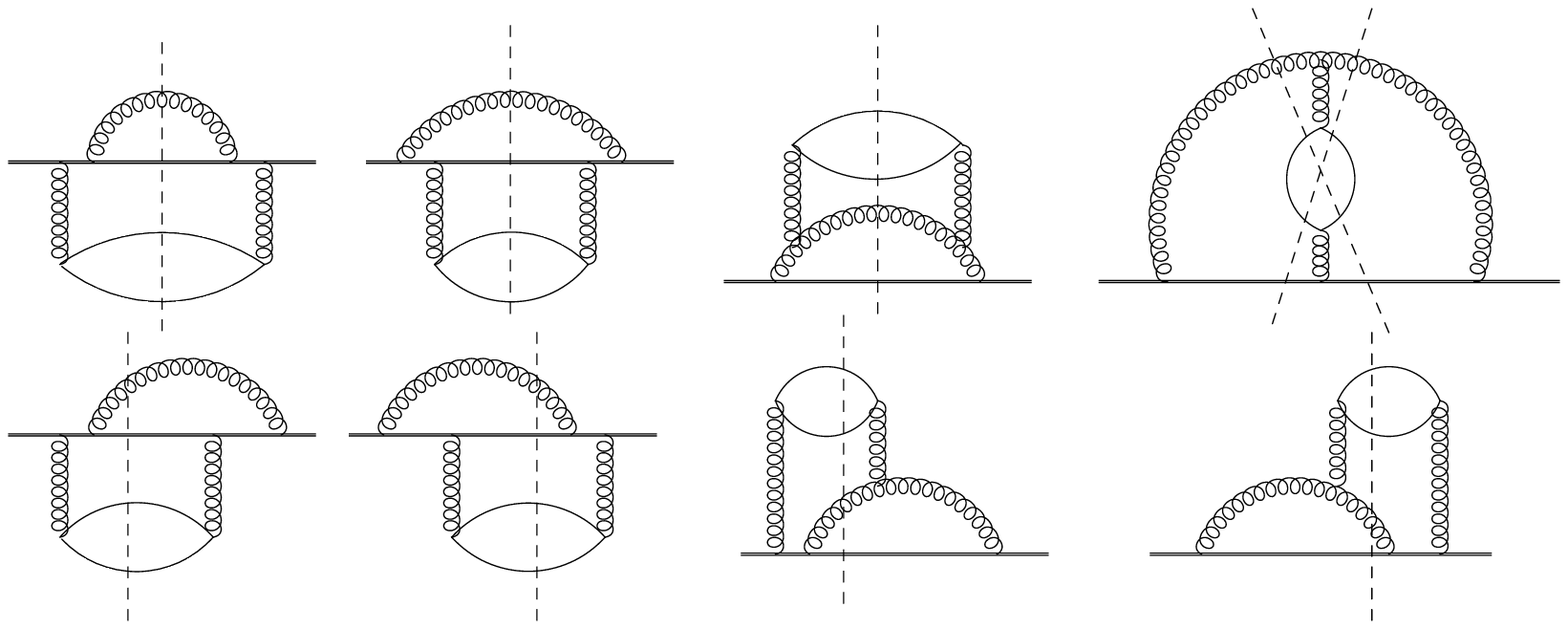}	
	\caption{The first row displays the square of the Born diagrams that can be generated from the cuts of the diagrams shown in the lower two rows.}
	\label{fig_borncut}
\end{figure}

\begin{figure} [!t]
	\centering
	\includegraphics[width=1\columnwidth]{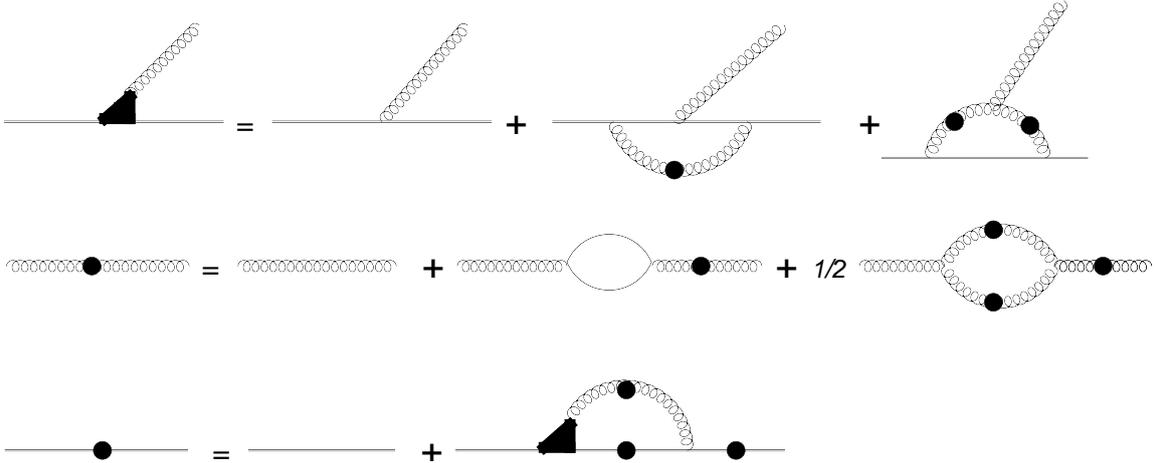}
	\caption{The coupled set of DSEs with skeleton diagrams that can generate the pQCD diagrams in Fig.~\ref{fig_borncut}.}
	\label{fig_dse}
\end{figure}

We start with the tree-level pQCD diagrams~\cite{Gunion:1981qs,Kang:2016ofv} shown in Fig.~\ref{fig_borncut}. The diagrams displayed in the first row represent the (naive) order-$g^2$ contributions to the amplitude for two incoming quarks going into two quarks and a gluon. The sum of these amplitudes is transverse ($ k^\mu M_\mu=0 $) and thus gauge invariant in the usual sense. 
%As discussed below Eq.~(\ref{eq_vetex}), although the individual piece of the diagram is not transverse or say gauge invariant, one can always project the transverse part of the diagram $M_{T\mu}= M_\nu P_T^{\nu\mu}$ so that one have $M_{T\mu}k^\mu=0$ and gauge invariance \footnote{$ P_T^{ij}=\delta^{ij}-k^ik^j/k^2,P_T^{0\mu}=P_T^{\mu0}=0$ \cite{Kapusta:2006pm}}. However, the problem for this procedure is that we will miss other gauge invariant piece at the same order. 
If we contract the external legs between terms in the left and right parentheses, we obtain the selfenergy diagrams in the second and third rows, which relate to the diagrams in the first row by cutting rules (optical theorem). 
%The selfenergy diagrams calculated in this scenario are also gauge invariant. 
As discussed in the previous section, the key quantities to evaluate the radiative processes are the 
selfenergies, which explicitly figure in Eq.~(\ref{KB}) or in the spectral functions as shown in Eq.~(\ref{eq_spec}). The NP selfenergies can be calculated through a series of coupled Dyson-Schwinger equations (DSEs); our approach falls into this category. The structure of the equations is controlled by the skeleton diagram expansion where an example is shown in Fig.~\ref{fig_dse}. In this example, by inserting the dressed vertex (first row) and the dressed gluon (second row) into the equation in third row, we can generate the pQCD selfenergy diagrams shown in second and third rows in Fig.~\ref{fig_borncut}.  With a specific choice of the skeleton expansion, the diagrams generated by DSEs can encompass any finite-order perturbative diagrams. In other words, using the DSEs perturbatively is an alternative and equivalent diagram expansion method for the perturbative diagram expansion. With a specific truncation scheme, the DSEs form a closed system of equations that allow NP solutions. These solutions can be obtained through selfconsistent iterations. They resum the diagrams with specific patterns (such as ladder diagrams, rainbow diagrams and ring diagrams) to infinite order, which goes beyond any fixed-order perturbative expansion and thus can provide meaningful results at large coupling strength. However, these truncation schemes usually compromise explicit gauge invariance since they select a particular subset of diagrams to resum. There is no general solution to this problem yet in the strongly interacting regime, although progress has been made in resolving this issue in vacuum~\cite{Roberts:1994dr,Roberts:2000aa,Maris:2003vk}.
	
In the present article we focus on NP features at large interaction strength. We compromise on exact gauge invariance by carrying out the $t$-channel ladder resummation but restrict ourselves to projecting out the gauge invariant part by applying the transverse projector,  $M_{T}^{\mu}=P_{T}^{\mu\nu} M_\nu $   \footnote{$ P_T^{ij}=\delta^{ij}-k^ik^j/k^2,P_T^{0\mu}=P_T^{\mu0}=0$ \cite{Kapusta:2006pm}}, \ie, we only keep the bare radiation vertex with emission of the gauge invariant transverse mode, while in principle also a gauge invariant longitudinal mode is available in medium~\cite{Kapusta:2006pm,Bellac:2011kqa}. Thereby we also neglect the vertex corrections which would be required in connection with an in-medium emission vertex. More rigorous methods are beyond the scope of this work.
	%{\red{Can we say something about the HQ expansion here?}}
	
%	\delete{To obtain gauge independent information from our gauge-dependent approach, we investigate our problem from 3 different angles. 	First, as we discussed in the previous section, we have checked 2 different gauges, which lead to modest numerical variations in our results. Second, we have checked the consequences of replacing the bare gluon emission vertex by an HTL vertex, which 	is constructed to render an in-medium pQCD calculations gauge invariant. {\red{Discuss illustrative results here}}.	Finally, within a fixed gauge (Coulomb gauge), we conduct calculations for four setups with different inputs representing different NP effects and compare their differences. Since they are all calculated in one gauge, the gauge artifacts will be similar in different setups so that differences between them contain gauge independent insights. The idea is that any gauge artifacts will not induce a large distortion on the relative effects in the fixed gauge, especially if they are much larger than variations between different gauges.} 
To highlight model-independent information obtained from our apporach, we conduct calculations for four setups with different inputs representing different NP effects and compare their differences. Since they are all calculated in one model, potential gauge artifacts are expected to be similar in the different setups so that the differences between them give insights into model-independent features. 
%\comment{I try to do some modification accordingly, but do we still need some discussion like this?}

%%%%%%%%%%%%%%%%%%%%%%%%%%%%%%%%%%%%%%%%%%%%%%%%%%%%%%%%%%%%%%%
\subsection{Four cases of NP effects}
\label{ssec_setup}
%%%%%%%%%%%%%%%%%%%%%%%%%%%%%%%%%%%%%%%%%%%%%%%%%%%%%%%%%%%%%%%
Within the formalism laid out above the NP effects that we study in this work are essentially encoded in the spectral functions (selfenergies) of Eq.~(\ref{eq_selfeels}) which figure in the emission rate, Eq.~(\ref{eq_ratedetail}). These effects are mainly from 3 sources: confining interactions, ladder resummations, and off-shell interactions, each of them playing a critical role in 
our many-body approach to heavy quarks in the QGP~\cite{Liu:2017qah}. In the following we will define 3 scenarios, in addition to the full results, which allow us to exhibit their effects on gluon radiation from a heavy quark traveling through the QGP. 

\begin{table}[H]
	\centering
	\resizebox{1\textwidth}{!}{  
		\begin{tabular}{|c|c|c|c|c|}\hline
			Cases&(1) $V_{C}$ Born &(2) $ V_{CS} $ Born&(3) T-matrix Onshell &(4) T-matrix Offshell \\\hline
			Interaction& Coulomb & Coulomb+String &Coulomb+String &  Coulomb+String\\\hline
			Resummation& 2nd-order& 2nd-order & All order&All order  \\\hline
			Medium& quasi-particle & quasi-particle & quasi-particle&off-shell spectra  \\\hline
	\end{tabular}}
	\caption{Four cases are labeled by numbers or abbreviations in the first row where their differences in interaction, resummation scheme, medium content are listed in the second, third, and fourth rows respectively.}
	\label{table_case}
\end{table}

\begin{figure} [H]
	\centering
	\includegraphics[width=1\columnwidth]{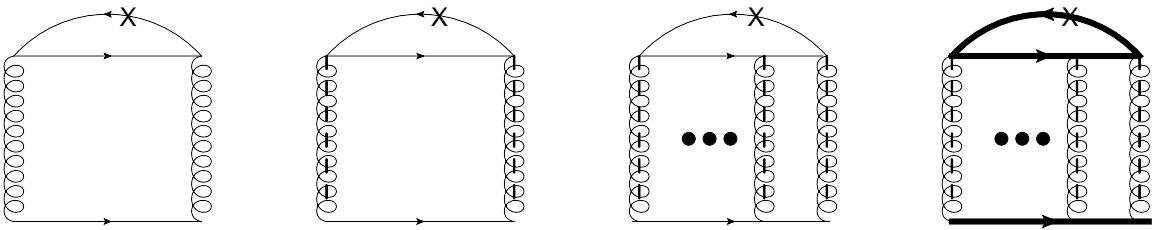}
	\caption{Self-energy diagrams for the four different cases discussed in Tab.~\ref{table_case}. Helical lines denote $V_C$, helical lines with double dashed lines denote $V_{CS}$, thin solid lines with arrows denote on-shell quasi-particle partons ($q$, $g$, HQ), thick solid lines with arrows denote off-shell dressed partons.}.
	\label{fig_sefe}
\end{figure}

The four cases are summarized in Tab.~\ref{table_case}, illustrated in Fig.~\ref{fig_sefe} and defined in detail in the following. 

Case-1, "$V_{C}$ Born", contains none of the three NP effects mentioned above, and as such represents the perturbative 
baseline in our approach. It only includes color-Coulomb interactions with coupling constant and Debye mass taken from the 
strongly coupled solution (SCS) in  Ref.~\cite{Liu:2018syc}. It does not contain a $t$-channel ladder resummation, and we only
keep the second-order Born diagrams as shown in Fig.~\ref{fig_sefe}, which are the leading-order diagrams that generate an
imaginary part of the selfenergy and consequently a finite width of the spectral functions of the outgoing heavy quark and gluon in Eq.(\ref{eq_ratedetail}) as required for a finite radiation rate. These diagrams are related to the amplitude squared of the Born scattering amplitude using the cutting rule. The thermal-medium partons are taken as zero-width quasiparticles with their masses 
fitted to the QGP EoS as discussed in our previous work~\cite{Liu:2018syc}. This case is most closely related to the treatment 
in typical pQCD calculations, although some differences persist, most notably the large thermal masses of the emitted gluon.

Case-2, "$V_{CS}$ Born" adds a NP effect to case-1, namely the confining interaction (potential) again with parameters taken 
from the SCS of Ref.~\cite{Liu:2018syc}, illustrated in the second panel by the extra bars in the gluon exchange lines. This is still done in leading-order Born approximation and with the same medium as in case-1. 

Case-3,  "T-matrix Onshell", additionally includes the $t$-channel ladder resummation in the heavy-light and light-light 
$T$-matrices (used to compute the outgoing HQ and gluon selfenergies) compared to case-2, which is illustrated in the third 
panel of Fig.~\ref{fig_sefe}. The other components are the same as in case (2). 

Finally, case-4, "T-matrix Offshell", uses finite width off-shell spectral functions for the thermal-medium partons. As shown in 
the last panel of Fig.~\ref{fig_sefe}, we now dress all internal lines in a selfconsistent caculation of selfenergies and $T$-matrices. 
In addition to the medium partons in the upper part of the diagram, the incoming and internal HQ lines are also dressed. The off-shell 
spectral functions are taken from the full solution of the SCS as discussed in Ref.~\cite{Liu:2017qah}. Case-4 is the most complete 
and consistent many-body theory calculation for the rate within our current approach, containing all three NP effects as highlighted 
at the beginning of this section.

Let us briefly reiterate on the role of the thermal-medium partons in the HQ transport calculation. For cases-1, -2 and -3, we use a zero-width quasiparticle medium. For case-4, the broad spectral functions that represent the off-shell medium are the predictions of the SCS in Ref.~\cite{Liu:2018syc}. Both scenarios describe the EoS of the QGP, implying that the effective density of "scattering centers" is quite similar, in order to focus on the genuine effects of the NP ingredients. Also note that the resummations carried 
out in the present work refer to $t$-channel ladder diagrams for a single (dynamic) scattering center. Higher orders due to multiple scattering centers and multiple gluon emission are implicitly included in the Langevin simulation, which is, of course, incoherent and thus neglects the Landau-Pomeranchuk-Migdal (LPM) effect~\cite{Landau:1953um,Migdal:1956tc,Gyulassy:1993hr}. While the latter is essential for radiation off light-flavor partons, it is mass suppressed for heavy quarks due to reduced formation times. A rigorous implementation of coherence effects in transport approaches remains challenging, see, \eg,  Ref.~\cite{Uphoff:2014hza} for an approximate treatment  in the HQ context. 

%Indeed, as we will show in a benchmark calculation in subsection~\ref{ssec_bench}, if we choose small parton masses, the rate of radiation will drastically enhanced. However, this density effects is not what we focused since we try to highlight the NP effects. 

%%%%%%%%%%%%%%%%%%%%%%%%%%%%%%%%%%%%%%%%%%%%%%%%%%%%%%%%
\section{Numerical Analysis} 
%Charm-Quark Transport Coefficients
\label{sec_num}
%%%%%%%%%%%%%%%%%%%%%%%%%%%%%%%%%%%%%%%%%%%%%%%%%%%%%%
In this section, we present and analyze the numerical results for the four cases defined above. We first study various microscopic quantities and then discuss how they manifest themselves in transport coefficients. We start by presenting the pertinent spectral functions for
heavy quarks and gluons in Sec.~\ref{ssec_spec}, followed by the corresponding power spectra in Sec.~\ref{ssec_power}. The resulting 
drag coefficients and fractional energy loss are calculated in Sec.~\ref{ssec_drag}. In Sec.~\ref{ssec_bench} we discuss our 
results in comparison to a pQCD calculation from the literature.

%%%%%%%%%%%%%%%%%%%%%%%%%%%%%%%%
\subsection{Spectral properties of radiated partons}
\label{ssec_spec}
\begin{figure} [t]
	\centering
	\includegraphics[width=1\columnwidth]{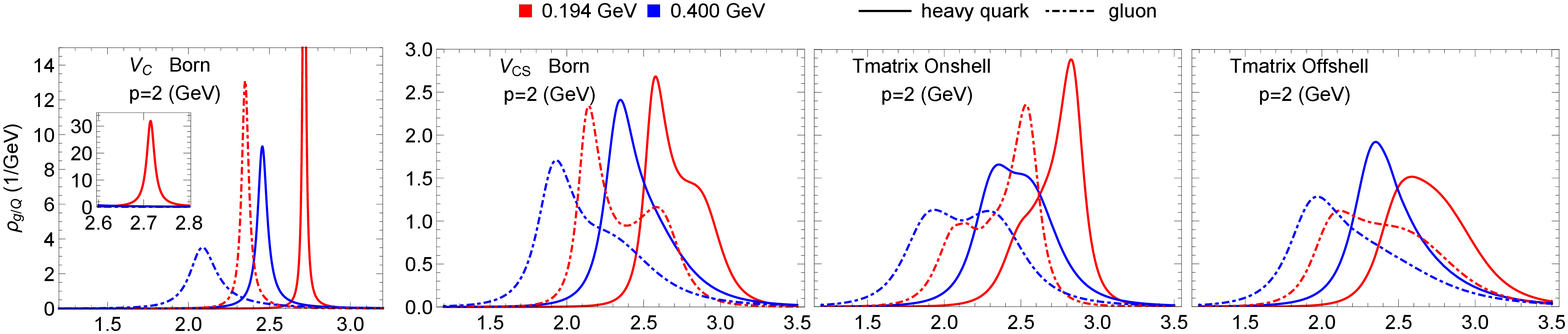}
	\includegraphics[width=1\columnwidth]{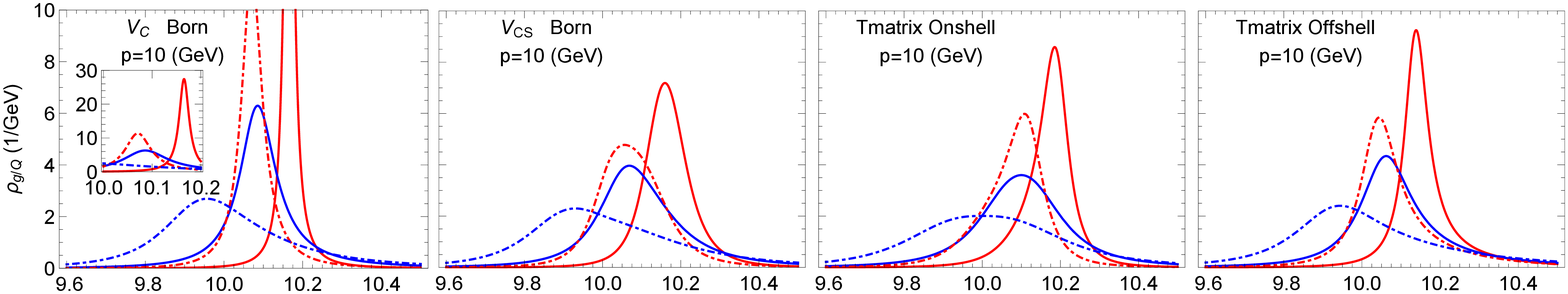}	
	\includegraphics[width=1\columnwidth]{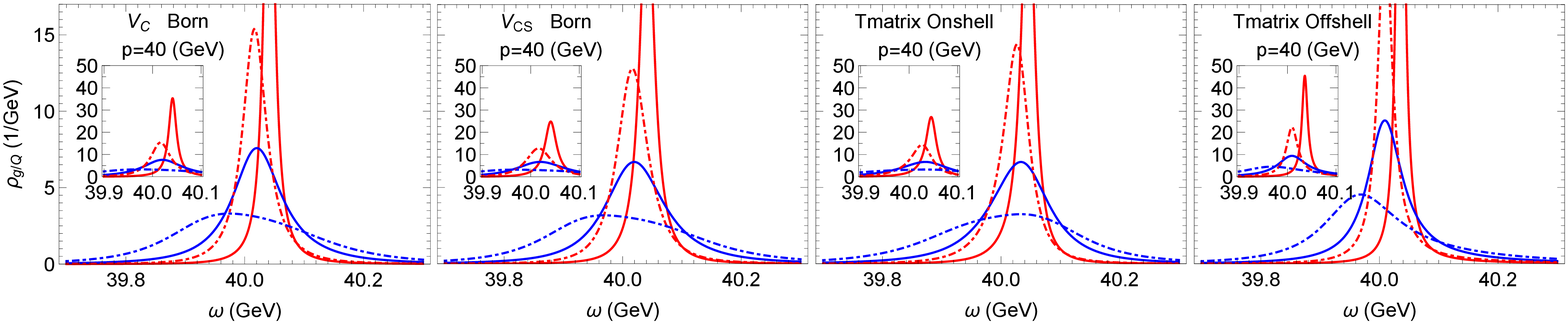}	\vspace{-0.cm}
	\caption{Spectral functions of charm quarks (solid lines) and gluons (dash-dotted) lines for cases 1 through from left to right , with different temperatures and momenta.}
	\label{fig_spec}
\end{figure}

%%%%%%%%%%%%%%%%%%%%%%%%%%%%%%%%
%As explained after Eq.~(\ref{eq_ratedetail}), the off-shell spectral function open up the phase space for simultaneously energy and momentum conservation in one to two splitting process, where its width (more accurately, the off-shell properties), encode the microscopic collisional process.
The spectral functions of the outgoing heavy quark and gluon are the key quantities for calculating the rate of radiation for
the $Q\to Qg$ process given by Eq.~(\ref{eq_ratedetail}). These spectral functions are shown in Fig.~\ref{fig_spec} for
our 4 different cases, each one for three HQ and gluon momenta, $p$=2, 10, 40~GeV and two temperatures , $T$=0.194, 0.4~GeV, as a function of energy around their on-shell values. We recall that for cases 1-3, these spectral functions differ from those in the internal lines in Fig.~\ref{fig_sefe} where they are zero-width quasiparticles. 
%We first focus on the features shared by most of these four cases.
%There are several general features of spectral functions. 
For all cases, a markedly different feature from the standard pQCD approach are the rather large thermal-parton masses dictated
by the constraints from the QGP EoS and microscopically related to the Polyakov loop encoding the nontrivial information of the confinement~\cite{Liu:2017qah} generated selfconsistently from the confining potential. Similar features from including effects of 
the Polyakov loop have recently been found from a different perspective, within the Polyakov-quark-meson 
model~\cite{Singh:2018wps}. 
%Also, the fitting to lattice EoS requires large masses. In addition, a large gluon mass is required to avoid the condensation when the interaction is strong. 
With a gluon mass of around 1~GeV at low temperature, soft radiation is heavily suppressed. 
%The widths of the spectral functions, which  result from the collision between partons, also share similar features on temperature and momentum dependence in most of the cases. 
With increasing 3-momentum at a fixed temperature (from top to bottom row in  Fig.~\ref{fig_spec}),  the spectral functions 
in the "$V_{C}$- Born" case 
(left panels) show an increasing width while the opposite trend is found for the other 3 cases which is caused by the presence of the 
confining interaction whose strength rapidly decreases with increasing momentum (transfer),  especially at low temperature. 
At the same time, at high momentum (bottom row), there is little difference among the different cases, with well defined 
quasiparticles characteristic for pQCD calculations. These features are a direct reflection of our previous findings that the 
QGP medium at moderate temperatures is strongly coupled at large wavelengths while recovering a more weakly interacting quasiparticle structure with increasing resolution~\cite{Liu:2017qah}. 
For the temperature dependence at a fixed 3-momentum, the "$V_{C}$ Born" case shows a clear increase in width from low to 
high temperatures, which is \add{a} consequence of the increasing thermal-parton density which overwhelms the moderate loss in 
interaction strength caused by a stronger Debye screening. This remains true for the other three cases at high momenta, while 
at low momentum this trend is much less pronounced, even slightly inverted for the "T-matrix Offshell" case. 
At the lowest temperature and momentum ($T=0.194$~GeV, $p=$2~GeV), all 3 cases (2-4) involving the confining interaction 
show appreciable distortions from a simple Lorentzian shape. In particular, the gluon spectral functions exhibit substantial
strength, even collective modes signaled by peaks,  in the spacelike region, \ie, for $\omega<p$  (for both temperatures), 
which is the most relevant one for the emission process of the incoming on-shell heavy quark. The spacelike strength tends to 
increase with the inclusion of more NP effects.
%Although the strength of the space-like region is correlated with the widths of the spectral function, in the cases where the spectral function are strongly deviated from a Lorentzian, the space-like region is better correlated with rates of radiations rather than the width. 
At high temperature and low momentum ($T=0.400$~GeV, $p=$2~GeV), the inclusion of the string interaction alone, at 
the Born level, strongly enhances the strength in the space-like region, compare the dash-dotted blue lines in the 2nd and 1st 
panel of the upper row in Fig.~\ref{fig_spec}. Although the addition of resummation (case-3) and off-shell (case-4) effects 
further distorts the spectral functions (cf.~3rd and 4th panel), they do not lead to significant differences for the phase space 
in the spacelike region. 
%Indeed, the formula~Eq.~(\ref{eq_ratedetail}) itself do allow emitting either a soft gluons or a soft HQ with high-energy gluons, where the space-like region of soft HQ/gluon is the responsible for soft HQ/gluon emission.
For higher momentum, $p$=10\,GeV, at both temperatures, the change of the spectral properties by adding the NP effects is less pronounced. Still, the largest modification is generated by the inclusion of the string interaction at low temperatures, while the 
further differences in cases 3 and 4 are not significant. At high temperature and $p$=10\,GeV, even the inclusion of the string 
interaction is no longer significant. 
%At the highest momentum $ p=40 $~GeV, for both temperatures, the change of the spectral properties by adding the NP effects are no longer significant. 
%We note the "T-matrix Offshell" with all NP effects have the smallest widths. This is because a full self-consistent evaluation usually has the tendency to push the pole of the propagator lower. The width (energy dependent) is smaller at pole  is smaller which leads to a smaller width {\color{red}(This sentence is strange)}.

%%%%%%%%%%%%%%%%%%%%%%
\subsection{Power spectra}
\label{ssec_power}
%%%%%%%%%%%%%%%%%%%%%
%The spectral properties discussed in subsection~\ref{ssec_spec} will emerged into the power spectra discussed in this subsection  
Inserting the spectral functions discussed in the previous section into the rate, Eq.~(\ref{eq_ratedetail}), we can evaluate the 
power spectra of radiation defined in Eqs.~(\ref{eq_power}) and (\ref{eq_xdnddx}). The former essentially corresponds the 
radiation spectrum, while the latter, with the transverse momentum transfer integrated, is more readily interpreted as the 
radiative energy loss of the heavy quark. At the end of this section we compute the transverse-momentum broadening 
coefficient, $\hat{q}$,  for the "$V_{C}$ Born'' and "$V_{CS}$ Born'' cases and illustrate how they relate back to their 
differences in the power spectra.

\begin{figure} [!t]
	\centering
	\includegraphics[width=1\columnwidth]{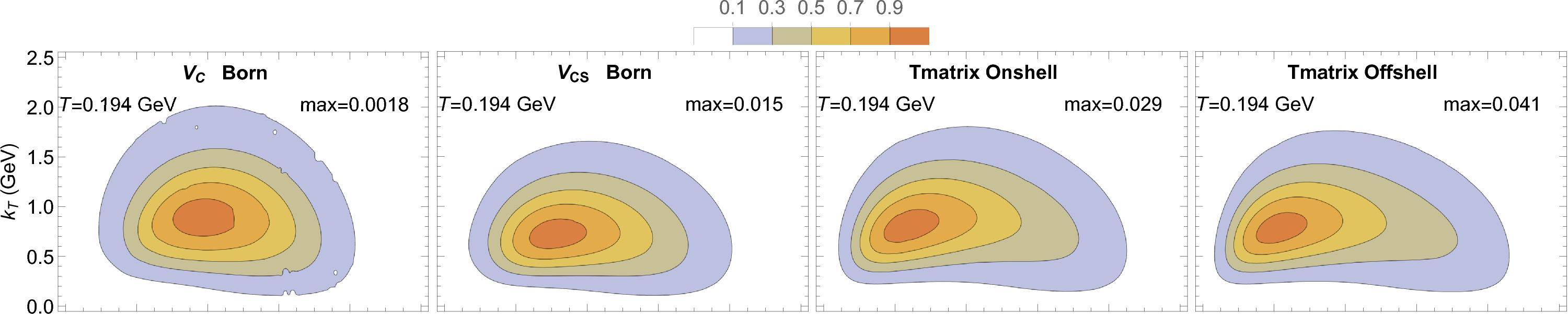}
	\includegraphics[width=1\columnwidth]{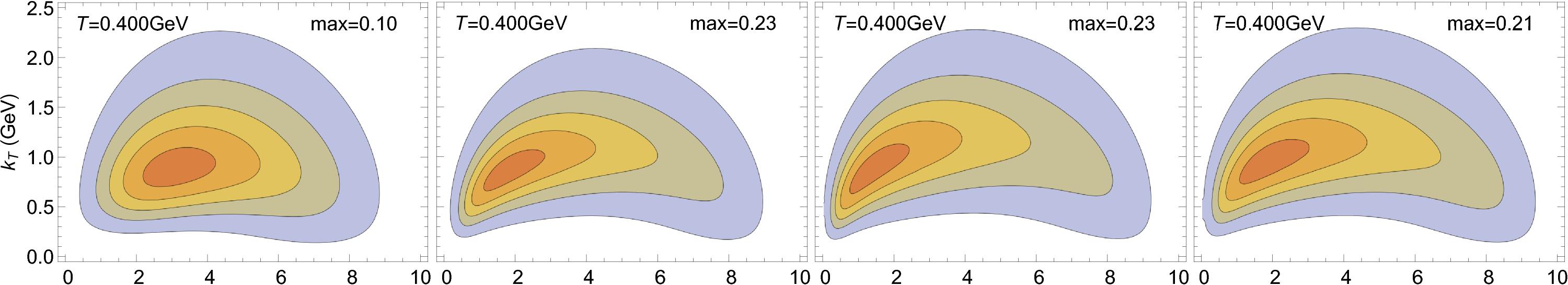}
	\includegraphics[width=1\columnwidth]{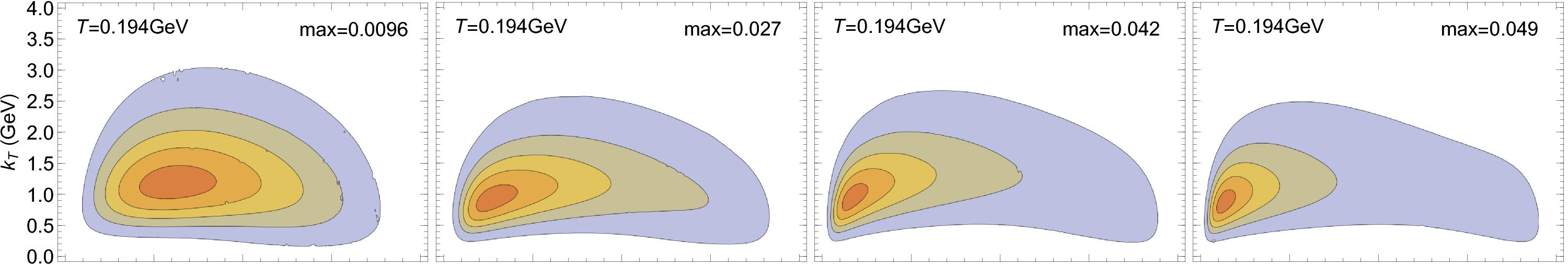}
	\includegraphics[width=1\columnwidth]{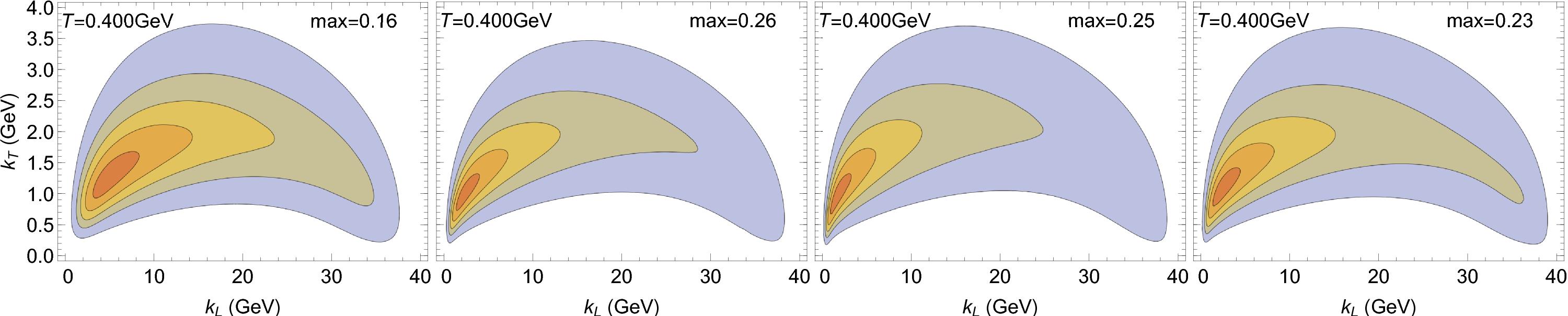}
	\caption{Contour plots of the normalized radiation spectrum, f $ W(k_T,k_L;p)/W_{max} $ in the plane of longitudinal 
($ k_L $) and transverse momentum ($ k_T $) of the emitted gluon for fixed HQ momentum and temperature. Each column corresponds to one of the four cases of NP effects, while the upper (lower) two rows are for $p$=10(40)\, GeV, and for different
temperatures ($T$=0.194(0.4)\,GeV for rows 1 and 3 (2 and 4)). The maximum values,  $W_{\rm max}$, of each the power spectra  are labeled as "max" with unit (GeV/fm) in the plots.}
	\label{fig_power}
\end{figure}

\begin{figure} [t]
	\centering
	\includegraphics[width=1\columnwidth]{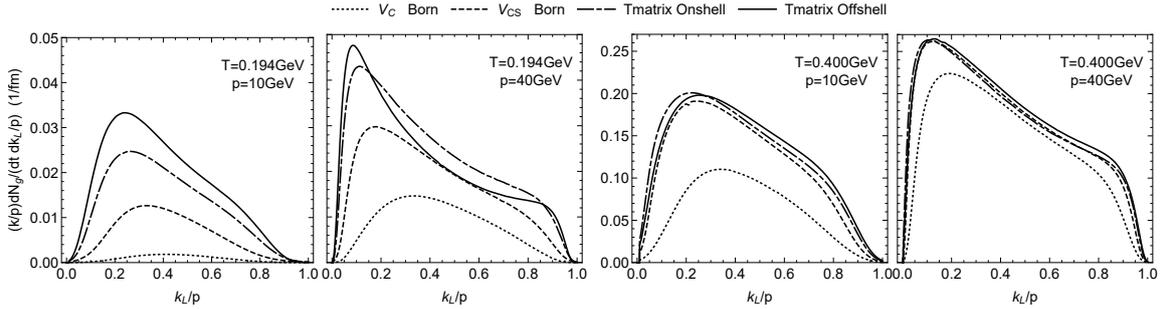}
	\caption{The $k_T$-integrated power spectra, $  \frac{(k/p)dN}{dt d(k_L/p)}\approx  \frac{xdN}{dt dx}$,  as a function 
of longitudinal-momentum fraction ($ k_L $) of the emitted gluon for different for the four different cases in each figure,
 at low (high) temperature in the left (right) two panels, each for two different HQ momenta.}
	\label{fig_powerin}
\end{figure}

In Fig.~\ref{fig_power} we display contour plots of the power spectra in the $k_T$-$k_L$ plane, where each of the 4 interaction cases is represented by each of the four columns, and each row represents a pair of temperature and HQ 3-momentum, for 
$p$=10~GeV with $T$=0.194, 0.4\,GeV in the upper 2 rows, and $p$=40\,GeV with $T$=0.194, 0.4\,GeV in the lower 2 rows.
The power spectra are projected onto the $k_L$ values (\ie, integrated over $k_T$) and shown in  Fig.~\ref{fig_powerin}, 
where the 4 combinations of $p$ and $T$ are combined into a single plot for each of our 4 cases. 
The generic features of the power spectra are a suppression of very soft radiation, \ie, for small $k_L$, due to the large effective gluon mass, and a  suppression of the collinear 
radiation, \ie, for small angles, $k_T/k_L$, which is the well known deadcone effect~\cite{Dokshitzer:2001zm}.
% -- can be observed in the power spectra as shown in Fig.~\ref{fig_power}, which indeed will be more obvious if one look at intensive (number density) spectra rather than power spectra as we checked. 
%On the other hand, the hard gluon radiation is typically quite collinear.
As expected, the radiation power increases with both temperature and momentum, roughly by one order of magnitude when going from $T$=0.194 to 0.4\,GeV,  and between a factor of 5 (for case-1 at T=0.194\,GeV) and 10\% (for cases 2-4 at $T$=0.4\,GeV) 
when going from $p$=10\,GeV to 40\,GeV (note that the contour plots have been scaled to their respective maxima in each panel, as quoted in the figure legend). The $k_T$ distributions of the power spectra tend to narrow down with increasing $p$.
% and also somewhat with increasing $T$}.
%The area of peak region tends to shrink and be more concentrated in low $k_L/p$ region as momentum increase  as shown  in Fig.~\ref{fig_power}.
%while they become slimmer as the HQ momentum increases as shown in Fig.~\ref{fig_power}.  
%This dead-cone seems more manifest at higher momentum   

The most signficant of the NP effects is the inclusion of the string interaction, when going from case-1 to case-2, and it is more pronounced for lower temperatures, as borne out from the maximum values in the power spectra which increase between a factor 
of $\sim$30 for ($p$,$T$)=(10,0.194)\,GeV and factor of $\sim$1.6 for  ($p$,$T$)=(40,0.4)\,GeV. 
% In general, the inclusion of more NP effects will allow stronger radiation, especially, stronger soft radiation. This is because the spectral functions at soft momentum have more strength in space-like regions from the NP effects (see earlier discussion on Fig.~\ref{fig_spec}). However, at high temperature ($ T $=0.400GeV,$p$=10~GeV), only the inclusion of the string interaction  strongly influences the power spectra, which can also be understood with the spectral properties discussed in Sec.~\ref{ssec_spec}. 
However, at low temperature, the additional NP effects in cases 3 and 4 (resummation and off-shell effects) are still rather significant,  especially for rather soft gluons ($x$$\ltsim$0.2), see the first and second panel in Fig.~\ref{fig_powerin}, while they 
have a small impact at high $T$, cf.~third and fourth panel in Fig.~\ref{fig_powerin}. These features directly reflect the 
discussion of the spectral properties of the outgoing gluon and heavy quark in the previous section.
%However, in soft-gluon region ($ x <0.2$ or $ k <8$~GeV) or soft quark region ($ x >0.8$ or $ k >$~32~GeV), difference by different NP effects are quite large due to the spectral properties at soft scales are quite different for all cases.
%This suggests the pQCD will be a good approximation to the full result, if the emitted gluon and quarks are quite hard. However, even at high temperature and high momentum, there are still significant differences between the "$V_{C}$ Born" and other cases in soft gluon or soft quark region, suggesting NP effects need to be considered in this region of the phase space, especially the effects of the confining force.
%if the emitted gluon or quark are soft, NP effects need to be considered, especially the effects of the confining force need to be included.
%
\begin{figure} [t]
	\centering
	\includegraphics[width=1\columnwidth]{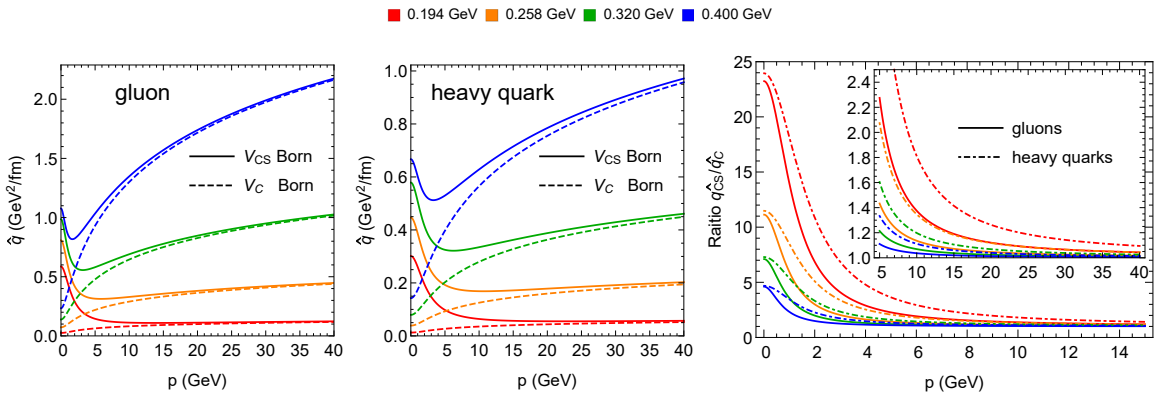}
	\caption{The transverse-momentum broadening coefficient, $ \hat{q} $, of a gluon (left panel) and a charm quark 
(moddle panel), calculated with a fixed coupling constant for different temperatures and HQ momenta in cases 1 (dashed
 lines) and 2 (solid lines). The right panel shows the ratios of the solid over dashed curves from the left and middle panel (solid curves for gluons and dashed curves for charm quarks). In all panels the red, orange, green and blue curves represent temperatures
$T$=0.194, 0.258, 0.32 and 0.4\,GeV, respectively.}
	\label{fig_qhat}
\end{figure}

\begin{figure} [h]
\hbox{  }
	\centering
	\includegraphics[width=0.6\columnwidth]{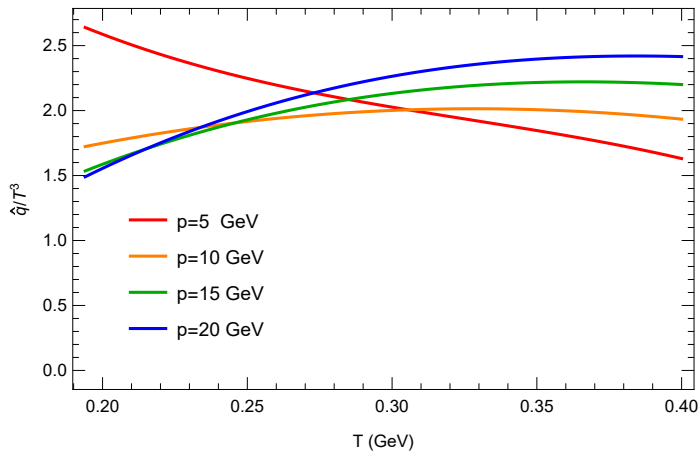}
	\caption{Dimensionless-scaled HQ transport coefficient, $\qhat/T^3$ , in the "$V_{CS}$ Born" scenario as a function 
of temperature for different HQ momenta.}
	\label{fig_qhat-T}
\end{figure}

The high-energy transport parameter $\hat{q}$ represents the average transverse-momentum transfer per mean-free-path 
of the heavy quark. It has been related to the elastic part of the commonly calculated coefficients in a Fokker-Planck 
equation~\cite{Svetitsky:1987gq}, specifically the transverse-momentum diffusion coefficient, $B_0^{\rm el}$, as 
$\hat{q}=4B^\text{el}_{0}\approx4T\epsilon_Q(p)A^{\text{el}}(p)$~\cite{Gubser:2006nz}; the second approximate equality involving the elastic friction coefficient, $A^{\text{el}}$, follows from the Einstein relation (which is routinely
enforced in practical applications).
In pQCD based approaches~\cite{Burke:2013yra}, $\hat{q}$ is usually factorized and used as a fit parameter encoding the 
NP properties of the QGP. Here we can check whether this factorization also holds at a more differential level, \eg, for the power spectra. Toward this end, we first compute the $\qhat$  for the "$V_{C}$ Born" and "$V_{CS}$ Born" scenarios 
from $ B^\text{el}_{0}$ with our Born amplitudes, to illustrate how the string interaction affects this widely used jet transport coefficient.
%(the other two cases require drastically more numerical efforts). 
The results are shown in Fig.~\ref{fig_qhat}. We here use a fixed coupling constant which implies that $\qhat$ increases 
logarithmically at high momentum, which is similar to the behavior found in Ref~\cite{Xu:2017obm} but different from 
results evaluated with a running coupling which typically come out approximately momentum independent. As 
expected from our preceding discussion of spectral functions and power spectra, the results for "$V_{CS}$ Born" and 
"$V_{C}$ Born" converge toward each other at high momentum, but the former is significantly enhanced for momenta 
below about 10\,GeV. This scale can thus be identified as the transition regime from perturbative  to the nonperturbative. 
A NP enhancement factor, defined in terms of the ratio of th $\qhat$'s from the 2 scenarios, is  shown in the right 
panel of Fig.~\ref{fig_qhat}; while the absolute value of the gluon $\qhat$ is larger than the one for charm quarks by 
about a factor of 2, the relative NP enhancement factor is actually larger for heavy quarks than for gluons. The 
temperature dependence of the HQ $\qhat$, scaled by $T^3$, is displayed in Fig.~\ref{fig_qhat-T} for 
the "$V_{CS}$ Born" case;  it reiterates the importance of the string interaction at relatively low momenta, significantly 
enhancing the coupling strength toward small temperatures.

\begin{figure} [t]
	\centering
	\includegraphics[width=1\columnwidth]{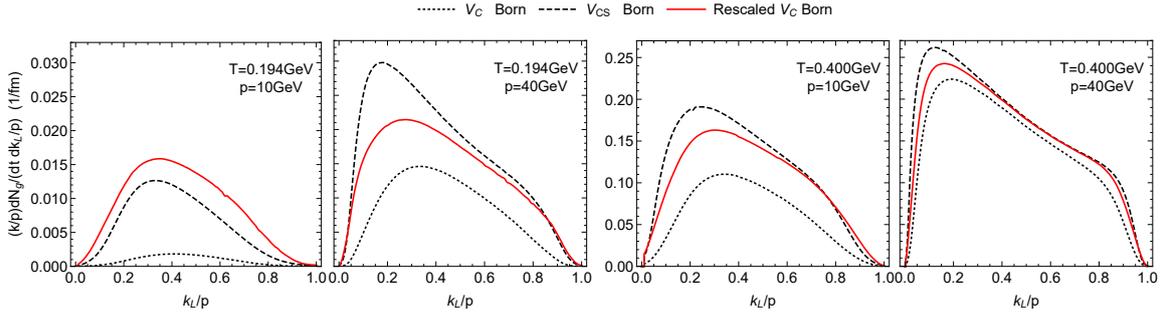}
	\caption{Comparison of the $k_T$-integrated power spectra$  \frac{(k/p)dN}{dt d(k_L/p)}\approx  \frac{xdN}{dt dx}$,  
as a function of longitudinal-momentum fraction ($ k_L $) of the emitted gluon for the "$V_{C}$ Born" case (dotted lines), 
the "$V_{CS}$ Born" case (dashed lines), and the "$V_{C}$ Born" case augmented with the $\qhat$  scaled enhancement factor 
from the two cases as defined in eq.~(\ref{qhat-scal}) (solid lines),  at low (high) temperature in the left (right) two panels, 
each for two different HQ momenta.}
	\label{fig_powerinrescale}
\end{figure}
We can the test the "factorization hypothesis" referred to above by multiplying the NP enhancement  factor into the power 
spectra of  "$V_{C}$ Born" and compare it to the power spectra of "$V_{CS}$ Born". More concretely, we define the 
"rescaled $V_{C}$ Born" power spectrum as 
\begin{align}
\left[\frac{(k/p)dN_g}{dt d(k_L/p)}\right]_\text{scaled}=\frac{(k/p)dN_g}{dt d(k_L/p)} \frac{\hat{q}_g^{CS}(k_L)}{\hat{q}_g^{C}(k_L)} \frac{\hat{q}_Q^{CS}(p-k_L)}{\hat{q}_Q^{C}(p-k_L)} \  ,
\label{qhat-scal}
\end{align} 
which accounts for the NP enhancement in both the emitted gluon and the outgoing heavy quark their respective emitted 
momenta. The results are plotted in Fig.~\ref{fig_powerinrescale} show that this prescription indeed gives an approximate 
mapping from "$V_{C}$ Born" to "$V_{CS}$ Born", although deviations of up to $\pm$30\% or so persist, especially at low temperature. 

\begin{figure} [h]
\hbox{  }
	\centering
	\includegraphics[width=0.6\columnwidth]{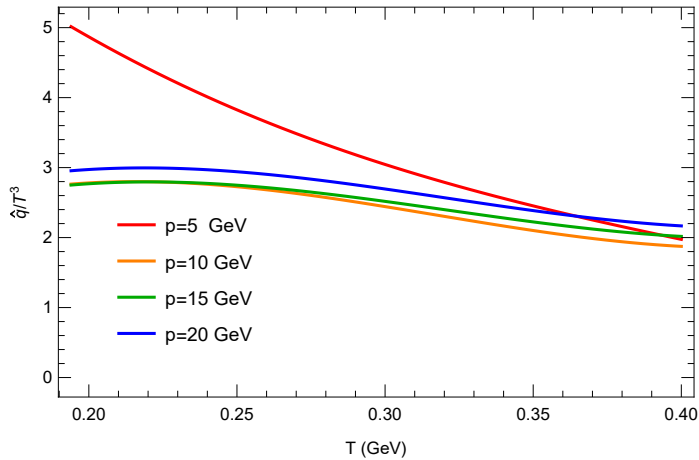}
	\caption{Dimensionless-scaled HQ transport coefficient, $\qhat/T^3$ , in the "$T$-matrix Offshell" scenario as a function 
of temperature for different HQ momenta.}
	\label{fig_qhat-T-full}
\end{figure}

For completeness, we display in Fig.~\ref{fig_qhat-T-full}  our full result for the temperature dependence of $\qhat$, \ie, in the 
"$T$-matrix Offshell" scenario, as obtained from the elastic friction coefficient, $A^{\rm el}(p)$, by use of the Einstein relation based on  Fig.~11 of Ref.~\cite{Liu:2018syc}  (which involves a partial-wave expansions up to $l$=8 to achieve a decent convergence at high momentum). It's magnitude is further enhanced by close to a factor of 2 at low momenta compared to the "$V_{CS}$ Born" scenario, while a significant temperature dependence mostly arises for momenta below 10\,GeV and temperatures below $\sim$300\,MeV.
 
%confining force drastically enhance the $\hat{q}$ at low momentum but  

%%%%%%%%%%%%%%%%%%%%%%%%%%%%%%
\subsection{Drag coefficient and energy loss}
\label{ssec_drag}
%%%%%%%%%%%%%%%%%%%%%%%%%%%%%%
The drag (or friction) coefficient, $A(p)$, is essentially an integral of the power spectrum over $ k_T $ and $ k_L $, divided by the HQ momentum, $ p $. It quantifies the fractional energy loss per unit time. Since for high-energy quarks the velocity is 
near the speed of light, it can be regarded as the fractional energy loss per unit path length. In the following, we discuss how the different cases affect the radiative drag coefficient, and implement them into a Langevin simulation for a charm quark in the 
background of QGP of finite size and fixed temperature, sometimes referred to as a "QGP brick". Several different energy loss 
approaches for charm and bottom quarks have been compared in this setup in Ref.~\cite{Rapp:2018qla}.

%employ this drag coefficients in a Langevin simulation in static medium to illustrate the energy loss for heavy quarks.
\begin{figure} [!b]
	\centering
	\includegraphics[width=1\columnwidth]{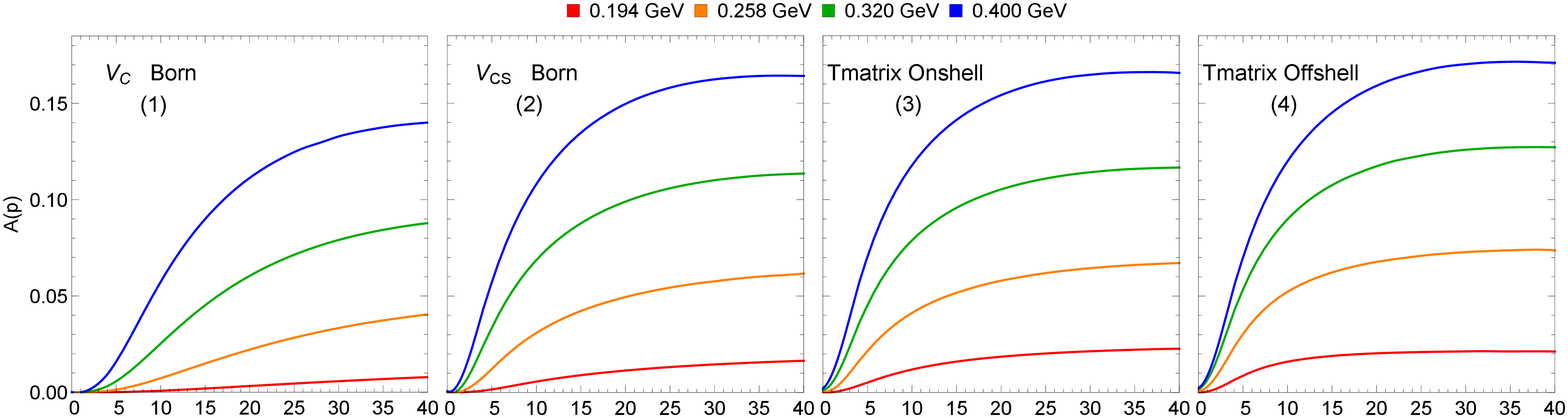}
	\includegraphics[width=1\columnwidth]{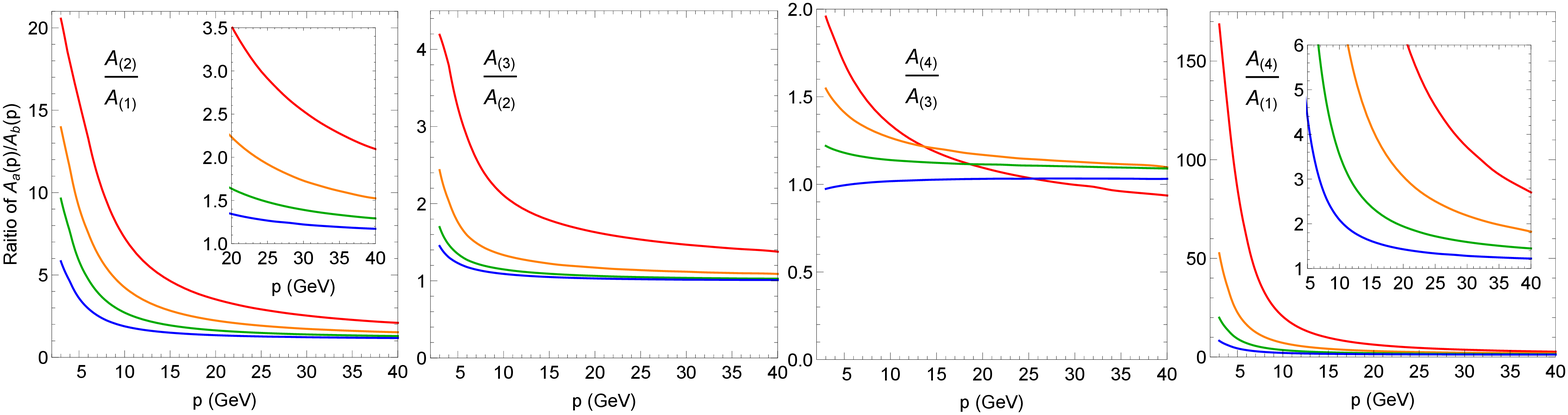}	
	\caption{The first row shows the drag coefficients for the four cases of NP effects, as a function of momentum for 
4 different tempertures each. In the second row the first three panels show the ratios $ A_{(i+1)}/A_i $ between the subsequent 
drag coefficients of the four cases, with the last panel showing the total effect, $A_4/A_1$.}
	\label{fig_drag}
\end{figure}
\begin{figure} [!t]
	\centering
	\includegraphics[width=1\columnwidth]{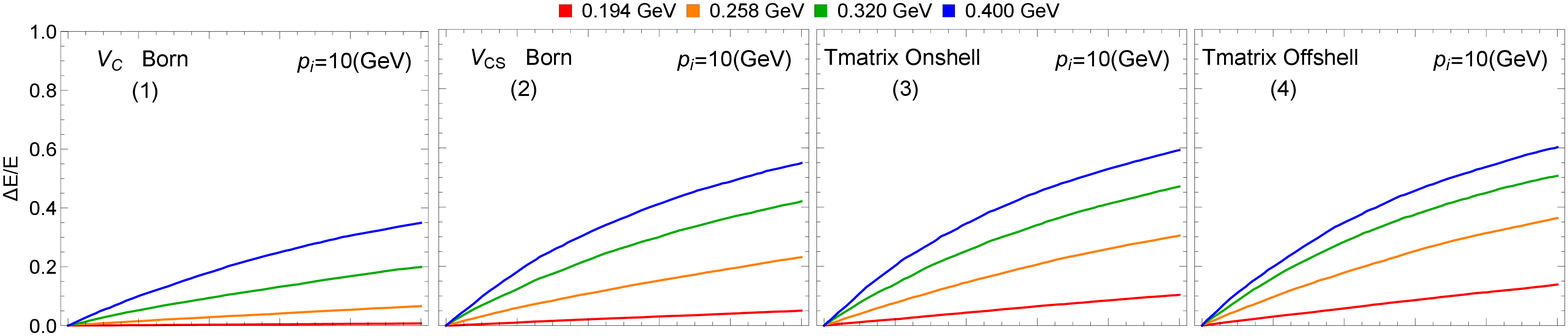}
	\includegraphics[width=1\columnwidth]{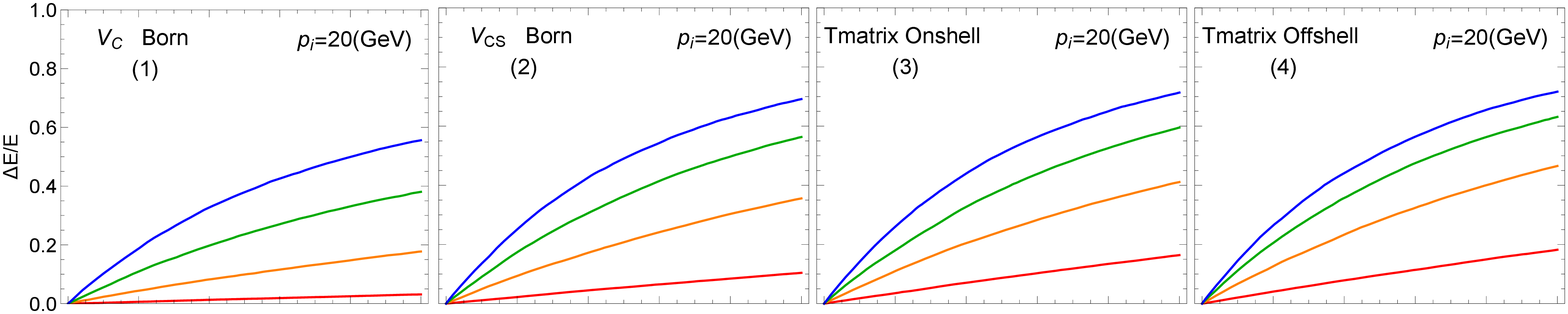}
	\includegraphics[width=1\columnwidth]{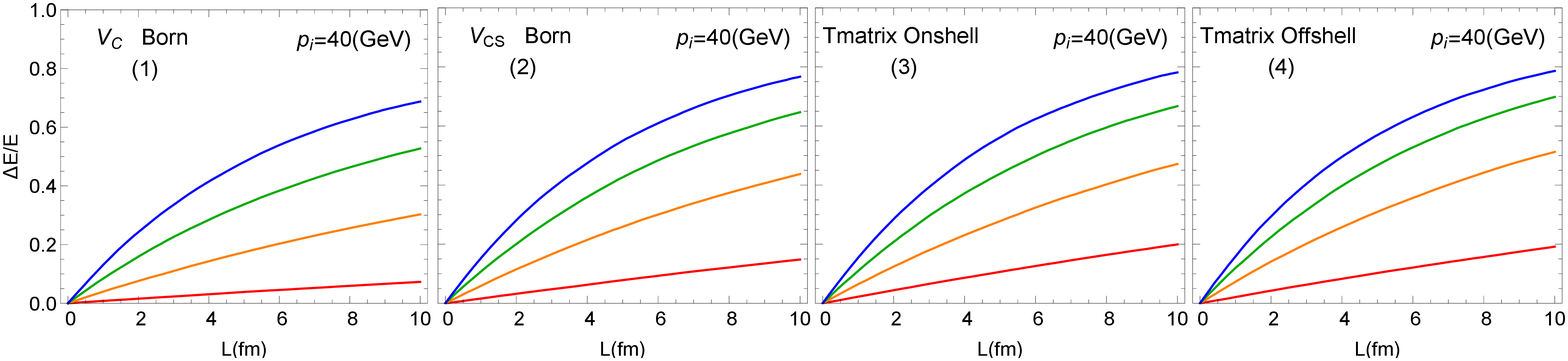}
	\caption{The fractional energy loss, $ \Delta E/E=(\varepsilon_Q(0)-\varepsilon_Q(L))/\varepsilon_Q(0)$, for different 
the 4 cases of NP effects, for initial HQ momenta of $ p_i=10,20,40$~GeV in the upper, middle and bottom row, respectively, 
and for 4 temperatures in each panel.}
	\label{fig_eloss}
\end{figure}

In the first row of Fig.~\ref{fig_drag}, we show the drag coefficient as a function of HQ momentum for the four different 
cases in each panel for different temperatures. At large momentum ($p$=40\,GeV), the drag coefficients essentially saturate 
at high $p$ and show a close-to-linear increase with temperature over the considered range,  $T$=0.194-0.4\,GeV, in all four cases; 
the latter is still true at lower momentum, $p$=10\,~GeV, except for the "$V_{C}$ Born" case which is closer to a quadratic 
temperature dependence. The drag coefficient is strongly suppressed at small momentum, as a consequence of the 
large thermal-gluon mass. 
%The saturation momentum scale tend to be larger at lower temperature, except the "Tmatrix Offshell" have a quite flat temperature dependence. Also, there is tendency that with more NP effects, the saturation momentum scale to be smaller at each temperature.
To better quantify the enhancement with the subsequent inclusion of NP effects we plot in the second row of  Fig.~\ref{fig_drag}  
the ratio between adjacent cases. The ratio $ A_{(2)}/A_{(1)} $ in the first panel,  "$V_{CS}$ Born" relative to "$V_{C}$ Born", 
shows the largest relative enhancement, most pronounced at lot $T$ and low $p$, as seen before,  reiterating the importance of 
the string interaction, even at the Born level. The ratio $ A_{(3)}/A_{(2)} $ in the second panel exhibits a more moderate, but still 
substantial  enhancement due to the resummation of $t$-channel ladder diagrams in the heavy-light $T$-matrix. The ratio 
$ A_{(4)}/A_{(3)} $ in the third panel, indicates that the additional off-shell treatment of the spectral  function  
of the medium partons has a still smaller impact; at high momentum, even the temperature hierarchy of this effect is not definite. 
The combined impact of the  NP effects on our pQCD baseline scenario, quantified by the ratio $ A_{(4)}/A_{(1)} $ shown in the lower right panel of Fig.~\ref{fig_drag}, enhances the radiative contribution to the drag coefficient by up to a factor of $\sim$150
at small momenta, below 3 \,GeV, and at temperatures close to $T_{\rm pc}$. This is, however, somewhat academic, since its absolute magnitude is very small compared to the elastic contribution in the same framework~\cite{Liu:2018syc}, \eg, 
$A_{\rm el}(p=3\,{\rm GeV},T=0.194\,{\rm GeV})=0.1/{\rm fm}$ vs. compared to 0.03/fm for the radiative drag. 
On the other hand, already at $p$=5\,GeV and $T$=0.4GeV, where both contributions are near 0.08/fm,  the NP enhancement 
of a factor of  $\sim$4 is certainly relevant. 

Finally, we implement the radiative transport coefficients into a Langevin simulation given
by incremental time steps for the HQ position and momentum, 
\begin{align}
&d\textbf{x}=\frac{\textbf{p}}{\varepsilon_c(p)} dt, \quad
d\textbf{p}=-\Gamma(p)\,\textbf{p}dt+\sqrt{2 dt D(p)}\bm{\rho}  \  ,
\label{eq_lan}
\end{align}
where $ \bm{\rho} $ is a random number determined from a Gaussian distribution function, 
$ P(\bm{\rho})=(2\pi)^{-3/2}e^{-\bm{\rho}/2} $, and the transport coefficients are  
\begin{align}
&\Gamma(p)=A(p),\quad D(p)=B_0(p)=B_1(p)=\int_p^\infty dq\, q \Gamma(q) e^{-\frac{\varepsilon_Q(q)-\varepsilon_Q(p)}{T}}\approx\Gamma(p) \varepsilon_Q(p) T\ .
\end{align}
We use the pre-point Ito setup in which the relaxation rate is equal to the drag coefficient, and the momentum 
diffusion coefficients are defined via the Einstein relation and obtained by solving the differential equation 
$ \Gamma(p)=1/\varepsilon_Q(D(p)/T-dD(p)/d\varepsilon_Q) $ with the boundary condition $ D(\infty)>0  $.
%
%where the relaxation rate, $\Gamma(p)=A(p)$, and the momentum diffusion coefficient, $D(p)$, $D(p)=B_0(p)=B_1(p)=\int_p^\infty dq \Gamma(q) \exp(-(\frac{\varepsilon_Q(q)-\varepsilon_Q(p)}{T}))T q$, and 
%$ \bm{\rho} $ is a random number determined from the Gaussian distribution function 
%$ P(\bm{\rho})=(2\pi)^{-3/2}e^{-\bm{\rho}/2} $. Using the Langevin equations, we simulate 
%Brownian motion of charm quarks in a background medium provided by an ideal hydrodynamic 
%evolution of the QGP fireball in URHICs at RHIC and the LHC. For definiteness, we choose 
%semicentral Pb-Pb collisions at CM energy $\sqrt{s_{\rm NN}}$=5.02\,TeV, at a fixed impact 
%parameter representing the 20-40\% centrality class. 
%
In Fig.~\ref{fig_eloss}, we show the results of 3D Langevin simulations for a charm quark inside a static QGP brick with 
a 10~fm path length. We have verified that the fluctuation term in Eq.~(\ref{eq_lan}) is not important for the average 
energy loss and therefore the Langevin equation can be replaced by a 1-D differential equation,
 $ dp=-\gamma(p) \varepsilon_Q(p)dx$. 
%This simulation in a static medium is a useful benchmark where similar calculations with other approaches can be found in Ref.~\cite{Rapp:2018qla}. 
Without the LPM effect~\cite{Gyulassy:1993hr}, the energy loss for small path lengths is near linear. For large path lengths, the
negative curvature is caused by the fact that the radiative drag coefficient decreases with decreasing HQ momentum. At low temperature and low momentum, our many-body approach generally predicts a much smaller radiation and energy loss than 
pQCD-based approaches (cf.~Fig.~27 in Ref.~\cite{Rapp:2018qla}). The main reason is the large gluon mass creating a large 
energy threshold for the emission. However, the thermal masses are a key feature of our many-body approach, essentially  
constrained  by the QGP equation of state, as elaborated  in Sec.~\ref{ssec_spec}. In other words, the emitted gluon 
can only propagate in the energy momentum modes supported by the ambient medium. This is quite different from approaches
where the emitted gluon in the QGP is assumed to be massless. The NP effects recover a good portion of the suppression in the
power spectra, as we will discuss in more detail in the following section. 
%On the other hand, for large initial HQ momentum and higher medium temperature, the energy loss predicted by our approach is quite similar to other approaches in  literature.

%%%%%%%%%%%%%%%%%%%%%%%%%%
\subsection{Comparison to a pQCD calculation}
\label{ssec_bench}
%%%%%%%%%%%%%%%%%%%%%%%%%%
\begin{figure} [t]
	\centering
	\includegraphics[width=0.7\columnwidth]{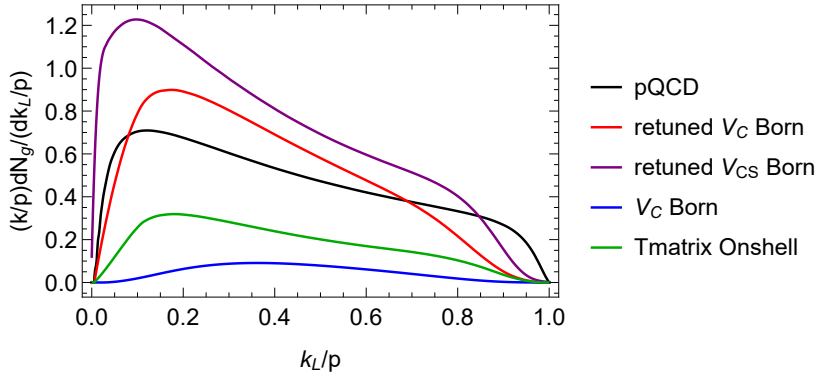}
	\caption{The quantity $\frac{(k/p)dN}{d(k_L/p)}\approx \frac{xdN}{dx}$ as a function of $k_L/p$. The initial HQ energy is 
20~GeV for a static medium of temperature $T$=225\,MeV and length $L$=5\,fm. The pQCD result is taken from 
Ref.~\cite{Djordjevic:2007at}.}
	\label{fig_powerpQCD}
\end{figure}
We finally carry out a smore quantitative comparison of our results to a  pQCD calculation from the 
literature~\cite{Djordjevic:2007at}. 
Toward this end we focus on the quantity $xdN/dx$ commonly displayed in pQCD approaches, which characterizes the energy loss
spectrum. We obtain this by integrating the power spectrum,  $xdN/dx=\int dt\,xdN/(dtdx)$, over time as obtained from
our Langevin simulation in Sec.~\ref{ssec_drag} which accounts for the time dependence of the charm-quark momentum,  $p(t)$.
In general, $xdN/dx$ is approximately equal to $xdN/(dtdx)$ multiplied by the length of the medium. We choose a QGP brick
of length $L$=5\,fm and temperature $T$=0.225\,GeV, and and initial  in charm-quark momentum of 20\,GeV.
%Fig.~\ref{fig_powerpQCD}, 
For a most direct comparison of our calculations to the pQCD approach, we evaluate our "$V_C$ Born" scenario with the same parameters as in Ref.~\cite{Djordjevic:2007at}, \ie, coupling constant $\alpha_s=0.3$, Coulomb Debye mass 
$m_d=g\sqrt{1+N_f/6}$ with $ N_f=2.5 $, thermal light-quark and gluon masses of $ m_q=m_d/\sqrt{6} $ and 
$m_g=m_d/\sqrt{2}$, respectively, and charm-quark mass $m_c=1.2$~GeV. At $T$=0.225~GeV, 
this amounts to  $m_d$=0.51~GeV, $m_q$=0.21~GeV, $m_g$=0.36~GeV and $m_c$=1.2~GeV. The result of this 
retuned case agrees with the pQCD result taken from Ref~\cite{Djordjevic:2007at} within $\sim$25\%, except for $x$ values 
close to 1 where our results go to zero faster, see Fig.~\ref{fig_powerpQCD}.
%\footnote{As checked, using massless light quarks and gluons can give us another $3\sim10$\% enhancement for most of region of x.} 
%We recall that we do not include LPM interference effects, which would presumably lead to an reduction of the 
%deviations at relatively small $x$.} \comment{the \cite{Djordjevic:2007at} also does not include LPM effects} 
This comparison gives us further confidence that contributions from gauge-invariant pieces that are missing in our 
approach are not large, and that the comparisons of the different NP effects 
(which are much larger) are meaningful. In particular, if we include in the retuned set-up the string interaction (in Born 
approximation), we find a rather significant enhancement over the Coulomb-only scenario, as before.
On the other hand, if we use our baseline parameters for the "$V_{C}$ Born" case ($\alpha_s=0.27$, 
$N_f$=3, $m_d$=0.42~GeV, $m_q$=0.43~GeV, $m_g$=1.15~GeV and $m_c$=1.76~GeV), the resulting energy loss 
spectrum is much reduced, cf.~Fig.~\ref{fig_powerpQCD}. While the interaction strength is quite similar to the retuned scenario,
the much larger quasiparticle masses fixed by fitting to the lQCD EoS reduce the color charge density and with it the rate of radiation. 
In addition, the larger gluon and charm-quark masses (the latter as constrained by the HQ free energy) suppress the phase 
space for radiation. If we include the NP effects, \ie, the string interactions and resummations, the energy loss spectrum
substantially inceases over the "$V_{C}$ Born" result but is still significantly below the pQCD calculation\footnote{As indicated in Sec.~\ref{ssec_power}, the drag coefficient from the "T-matrix Onshell"  case is rather close to the "T-matrix Offshell" case, 
while the latter is computationally much more intensive.}. The large interaction strength cannot compensate the loss caused by 
the large NP masses, which are largely generated by the (remnants of the) confining force in our approach. 
%This significant reduction of the radiation strength compared to the pQCD approach at relatively low momentum and low temperature is a feature of our approach, which may lead to measurable phenomena in future experiments.
%Although the final results for radiation is similar,  this microscopic picture is quite different. Our approaches is driven by strong interaction strength and other NP effects, while the pQCD in Ref~\cite{Djordjevic:2007at} is mostly driven by large color charge density.
%However, some pQCD framework also partly includes some essential NP effects. For example, implementing $\hat{q}$ in high twist approaches~\cite{Majumder:2010qh} can effectively include these NP effects, which we have discussed in more details in subsection~\ref{ssec_power}.

%%%%%%%%%%%%
\section{Conclusion}
\label{sec_con}
%%%%%%%%%%%%
%\red{we prospose the relatively large radiation is due to strong interaction rather than light parton mass}
We have analyzed the radiative energy loss of heavy quarks within a thermodynamic $T$-matrix approach, on the same 
footing with earlier studies of collisional energy loss. The most relevant aspects of this calculation are its nonperturbative 
components, specifically remnants of the confining force above $T_{\rm pc}$,  $t$-channel ladder resummations and 
off-shell spectral functions. These have been previously constrained by various sets of lattice-QCD data, most notably the 
QGP equation of state and heavy-quark free energy. To scrutinize their relevance, we have set up four scenarios with a 
subsequently increasing degree of NP effects.
 
We have found that all three NP effects referred to above substantially affect the power spectra and radiative transport coefficients at soft energy scales, \ie, at low HQ and/or gluon momenta and low temperatures. Ranking them by their 
importance, the inclusion of string interactions (even in Born approximation) generates the largest enhancement, followed by 
$t$-channel resummations in the two-body $T$-matrix, while the off-shell medium induces comparatively small modifications. 
Furthermore, we explicitly showed  that the NP effects become gradually suppressed with increasing resolution scale; \eg, 
at $T$=0.4\,GeV the NP enhancement in the radiative drag coefficient  amounts to less than 40(20)\% for charm quarks of 
momenta of $\sim$20(40)\,GeV.
% This hard scale can be achieved by increasing momentum or temperature, or focusing on the phase space of hard radiation. 
%Indeed, the $\alpha_s\sim0.3$ is not very small for perturbative expansions. 
This supports the convergence of pQCD-based approaches at high parton energies. 
While our calculations are carried out in a fixed gauge, we have checked that, when using color-Coulomb Born interactions 
and matching the input parameters to state-of-the-art pQCD calculations, our results for energy loss spectra for a high-energy 
charm quark agree with the latter within $\sim$25\% or so. However,
when using the thermal parton masses as dictated by the constraints from lattice-QCD, our baseline "pQCD" calculation results in
a strongly suppressed radiation spectrum, mostly due to the large energy cost (mass) of the thermal modes available 
to the radiated gluons in the ambient QGP medium. The combined enhancement effect of the NP interactions cannot fully 
recover this suppression. It will thus be important to further srutinize the robustness of this result, and investigate its impact on in the phenomenology of HF observables in URHICs.

%It is important to keep in mind, though, that That being said, an essential ortant feature of our approach is the large masses, especially the large gluon mass can suppress the radiation at low momentum scale, which can probably be tested with the future high precision measurements of single hadrons, jets and heavy-quarks over a wide energy range. For the future development, we will include the LPM effects in our approach which can help extend our approach to light partons. Furthermore, a simulation in a realistic QGP medium can help understand how this drag coefficient affects the D meson spectra observed in experiments.

\acknowledgments
We are grateful to helpful discussions with Shanshan Cao and Yi-Lun Du. 
This work has been supported by the U.S.~National Science Foundation (NSF) through grant PHY-1913286.

\appendix

\bibliographystyle{JHEP}
\bibliography{refcnew}

\end{document}